\documentclass[prd,twocolumn,showpacs,preprintnumbers,floats,floatfix,
amssymb,amsmath,nofootinbib]{revtex4}
\usepackage{graphicx}
\usepackage{dcolumn}
\usepackage{bm}
\def\spose#1{\hbox to 0pt{#1\hss}}
\def\lta{\mathrel{\spose{\lower 3pt\hbox{$\mathchar"218$}}
     \raise 2.0pt\hbox{$\mathchar"13C$}}}
\def\gta{\mathrel{\spose{\lower 3pt\hbox{$\mathchar"218$}}
     \raise 2.0pt\hbox{$\mathchar"13E$}}}
\newcommand{\be}{\begin{equation}}
\newcommand{\en}{\end{equation}}
\newcommand{\bea}{\begin{eqnarray}}
\newcommand{\ena}{\end{eqnarray}}
\newcommand{\ex}{\mbox{e}}
\newcommand{\dd}{\mbox{d}}
\newcommand{\cS}{c_{_{\rm S}}}
\newcommand{\cs}{c_{_{\rm S}}}
\newcommand{\nS}{n_{_{\rm S}}}
\newcommand{\mP}{m_{_{\rm Pl}}}
\newcommand{\zBST}{\zeta_{_{\mathrm{BST}}}}

\newcommand{\Hu}{{\cal H}} \newcommand{\Ka}{{\cal K}}

\begin{document}

\title{Passing through the bounce in the ekpyrotic models}

\author{J\'er\^ome Martin and Patrick Peter}
\email{jmartin@iap.fr, peter@iap.fr} \affiliation{Institut
d'Astrophysique de Paris, UPR 341, CNRS, 98 boulevard Arago, 
75014 Paris, France}
\author{Nelson Pinto-Neto}
\email{nelsonpn@cbpf.br} \affiliation{Centro Brasileiro de
Pesquisas F\'\i sicas, Rua Dr.  Xavier Sigaud 150, Urca 22290-180, 
Rio de Janeiro, RJ, Brazil}
\author{Dominik J.~Schwarz}
\email{dschwarz@hep.itp.tuwien.ac.at}
\affiliation{Institut f\"ur Theoretische Physik, 
             Technische Universit\"at Wien,
             Wiedner Hauptstra\ss e 8--10, 1040 Wien, Austria} 

\date{\today}
\begin{abstract}
By considering a simplified but exact model for realizing the
ekpyrotic scenario, we clarify various assumptions that have been used
in the literature. In particular, we discuss the new ekpyrotic
prescription for passing the perturbations through the singularity
which we show to provide a spectrum depending on a non physical
normalization function. We also show that this prescription does not
reproduce the exact result for a sharp transition. Then, more
generally, we demonstrate that, in the only case where a bounce can be
obtained in Einstein General Relativity without facing singularities
and/or violation of the standard energy conditions, the bounce cannot
be made arbitrarily short. This contrasts with the standard
(inflationary) situation where the transition between two eras with
different values of the equation of state can be considered as
instantaneous. We then argue that the usually conserved quantities are
not constant on a typical bounce time scale.  Finally, we also examine
the case of a test scalar field (or gravitational waves) where similar
results are obtained. We conclude that the full dynamical equations of
the underlying theory should be solved in a non singular case before
any conclusion can be drawn.

\end{abstract}
 \maketitle

\section{Introduction}

Modern ideas of particle physics, such as superstring,
$M-$theory~\cite{sugrastring} or quantum gravity~\cite{QG}, cannot in
general be subject to experimental constraints because of the enormous
energies (usually of the order of the Planck mass) at which they are
supposed to become effective. According to recent theoretical
developments~\cite{largeD,RS}, there is hope that space possesses more
than three large dimensions and that these extra dimensions might turn
out to be observable in a not too distant future. The majority of the
theoretical models that have been built so far are however based
on extremely high energy extensions of the standard particle physics
model, and thus currently need to be tested by the yardstick of
cosmology, the latter being the only playground at which those
theories could have acted.

According to the now standard paradigm that describes the early
universe and that is expected to stem from such high energy
particle models, a phase of superluminal accelerated expansion
known as inflation~\cite{inflation} preceded the
radiation-dominated epoch. Up to now, no model has come as close to
being a reasonable challenger to solve the standard cosmological
puzzles (flatness, homogeneity and monopole excess). The extra
bonus provided by the inflationary phase is that it leads
naturally to a scale-invariant density fluctuation spectrum that
seems to be in agreement with the observations.

Inspired by the recent developments of $M-$theory~\cite{heterotic}, in
particular through Ref.~\cite{HW}, and invoking brane cosmology,
recent work~\cite{ekp,ekpN,perturbekp} claimed to be able to solve all
the aforementioned problems as well, including a new way of producing
primordial cosmological perturbations. Although the model, both in its
``old''~\cite{ekp} and ``new''~\cite{ekpN} versions is plagued with
many difficulties~\cite{pyro}, as a potential alternative to inflation
(see also Ref.~\cite{PBB} in that respect), it is worth examining in
detail, would it be only to re-enforce the confidence we may have in
the latter.

In both the original and most recent versions, the
universe is supposed to consist of a four dimensional (visible) brane
evolving in a higher (in practice 5) dimensional bulk. By assuming the
brane to be a Bogomolnyi-Prasad-Sommerfield (BPS) state~\cite{BPS},
one ensures that the curvature $\Ka$ vanishes, thus addressing the
flatness problem. To begin with, another brane, that can be either a
light bulk brane~\cite{ekp}, or the other (hidden) boundary
brane~\cite{ekpN}, moves freely in the bulk until it collides with the
visible brane. The collision time is interpreted as the hot big bang
at which point the model is made to match the standard cosmological
model.

Apart from the collision time, the theory, which can be seen as
effectively four dimensional in the long wavelength limit, relies on
the General Relativity (GR) theory together with some extra fields. In
this effective 4D model, the Universe collapses, experiences a bounce
at some instant in time, and starts expanding. As far as cosmological
perturbations are concerned, only GR calculations have been discussed
up to now.

The pre-impact phase has been the subject of many tentative
calculations of the perturbation spectrum that would be generated by
quantum perturbations of the brane~\cite{perturbekp}. A general
agreement has now been reached~\cite{perturbekp,Lyth,BF,Hwang} that
the curvature perturbation spectrum $P_{\zeta }$ has spectral index
$n_\zeta=3$, while that of the Bardeen potential $P_{\rm \Phi }$ ends
up with $\nS=1$, i.e., a scale invariant spectrum. On the other hand,
the spectra of $\Phi$ and $\zeta$ are identical in the post-impact
phase, and enter the Cosmic Microwave Background Radiation (CMBR)
multipole moments.  It is therefore of utmost importance to obtain
full knowledge of these spectra not only in the pre-impact phase, but
also after the bounce has occurred, i.e., at times that are observable
now. In other words, the fate of $\Phi$ and $\zeta$ through the bounce
is the main issue before any conclusion regarding the model can be
drawn.

Only a few definite statements can be done about the bounce epoch. The
first, which was advocated by many authors, is that GR does hold
during it, or, stated differently, that it lasts sufficiently little
that corrections to GR can be regarded as negligible. Lacking the
actual theory, this is the only statement that can be endowed with a
predictive power. To begin with, it implies that there was no
singularity, and, if the null energy condition is to be satisfied,
that space is positively curve, i.e., $\Ka=1$. Under these conditions,
ordinary perturbation theory~\cite{Bardeen,perturb} can be applied. By
assuming continuity of the Bardeen potential and the well known
conserved quantity $\zeta$ (defined below), it was then
found~\cite{BF} that the scale invariant spectrum does not survive the
bounce, with the actual resulting spectrum being much lower than the
observed one. The temporary conclusion of this fact is that in order
that the ekpyrotic model be still compatible with the observation, a
new procedure must be applied to the bounce.

Arguing against GR during the bounce epoch sounds natural, as in
particular either the real theory is at least 5-dimensional, or, worst
indeed, in the case of the new scenario~\cite{ekpN}, the manifold
becomes (curvature) singular there, obviously leading to a breakdown
of ordinary GR across the bounce. In this case, a new criterion should
be derived to replace the ordinary junction conditions. Such a
criterion was provided in Ref.~\cite{perturbekp}, although without a
physical motivation, leading to the recovery of the observationally
correct spectrum. The very exhibition of junction conditions leading
to a scale invariant spectrum could then be seen as a hint that
constructing a realistic theory satisfying observational constraints
was not impossible.

Even if one is prepared to accept such drastic changes in the standard
cosmological picture, one might wonder as to the use of perturbation
theory on top of an otherwise singular
background~\cite{Lyth}. Moreover, it should be mentioned that
although the old scenario, because describable as an effective bounce
occurring at a low enough temperature, was avoiding the
over-production of grand unified scales monopoles~\cite{monop}, the
new model, being singular, poses this problem in a way which is as
acute as it was before the advent of inflation. Finally, the puzzle of
trans-Planckian scales~\cite{transPl}, quoted in Ref.~\cite{reply} as
a caveat for inflation, can be transposed in the new ekpyrotic model
in the same words.

This article is organized as follows. After a brief reminder of the
ekpyrotic model of the universe (Sec.~\ref{sec:ekp}), we examine in
detail the junction conditions suggested in Ref.~\cite{perturbekp}
(Sec.~\ref{sec:pert}). We concentrate in particular on the fact that
this proposed criterion rests on an altogether arbitrary (hence
unphysical) normalization function, so that whatever spectrum can be
obtained: obtaining a scale invariant spectrum in this model thus
turns out to be equivalent to imposing it from the outset. We also
demonstrate that the new prescription leads to an incorrect prediction
in the exact case of a radiation to matter domination transition.

We then consider a second possibility, i.e., we examine an effective
bounce in a context where the linear perturbation theory is 
still valid. We therefore considered first, in
section~\ref{sec:hydro}, the simplest case in which not only does GR
apply, but also in which all the calculations can be performed
analytically and consistently (indeed providing a nice textbook example for
cosmological perturbation theory illustration), namely that of a
${\cal K}=1$ bouncing universe with hydrodynamic 
perturbations~\cite{ppnpn1}. Then, using this toy model, we examine how 
the relevant perturbed quantities behave through the bounce. We 
pay special attention to the ``short time bounce limit'' (this is 
related to the question ``how sharp is sharp'' evoked 
in Ref.~\cite{Lyth}) and study whether, in this limit, the 
bounce can be considered as a 
surface where the equation of state jumps. If so this would 
allow us to use the standard junction conditions.

The second example that one can treat completely is by considering a
test scalar field. Indeed, in this case, one does not need to specify
what the origin of the background evolution is. In
section~\ref{sec:scalar}, we calculate the spectrum of a spectator
scalar field in such a bouncing background. Assuming no strong
deviation from GR at the perturbed level (we remind that such
deviations are necessary in the bounce region), and $\Ka=0$, this also
gives the tensor perturbation spectrum. The description of a bouncing
universe with ${\cal K}=0$ requires special care, as GR does not allow
for such a configuration to take place unless the Null Energy
Condition (NEC) is violated. Although this case is clearly contrived,
it provides at least an example where some arguments presented
recently in the literature can be implemented concretely, at the level
of equations.

\section{The Ekpyrotic scenarios}
\label{sec:ekp}

The ekpyrotic model is supposed~\cite{ekp,ekpN} to stem from the
theory by Ho\v{r}ava and Witten~\cite{HW} and some particular
construction of heterotic M$-$theory~\cite{heterotic}. It finds its
inspiration in the extra dimensional scenarios, {\it \`a la}
Randall~--~Sundrum~\cite{RS}, and can be motivated by compactifying
the action of 11 dimensional supergravity on an $S^1/Z_2$ orbifold,
compactified on a Calabi--Yau three-fold.  This results in an
effectively five dimensional action reading
\begin{equation}
{\cal S}_{_5} \propto \int_{{\cal M}_5} \dd^5x \sqrt{-g_{_5}}
\left[ R_{_{(5)}} -{1\over 2} \left(\partial\varphi\right)^2
-{3\over 2} { \ex^{2\varphi}{\cal F}^2\over 5~!}\right],
\end{equation}
where $\phi$ is the scalar modulus, and ${\cal F}$ the field
strength of a four-form gauge field. Two four--dimensional
boundary branes (orbifold fixed planes), one of which to be later
identified with our universe, are separated by a finite gap. Both
are BPS states~\cite{BPS}, i.e., they can be described at low
energy by an effective $N=1$ supersymmetric model, so that their
curvature vanishes. This is how the flatness problem is addressed
in the ekpyrotic model.

\begin{figure}[ht]
\includegraphics*[width=7cm, angle=0]{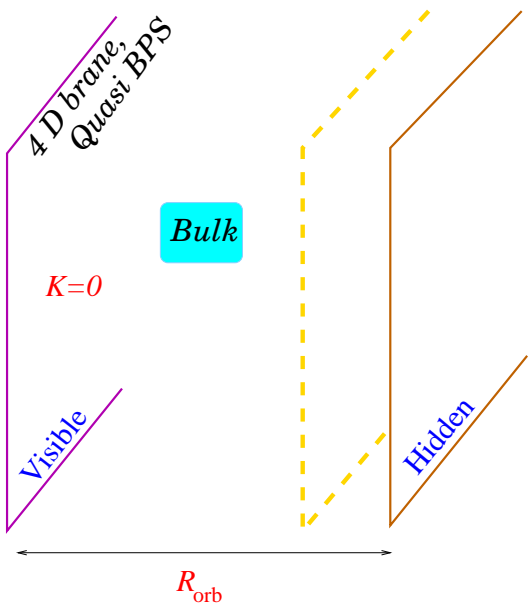}
\caption{Schematic representation of the old ekpyrotic model as a
bulk~--~boundary branes in an effective five dimensional theory.
Our Universe is to be identified with the visible brane, and a
bulk brane is spontaneously nucleated near the hidden brane,
moving towards our universe to produce the Big-Bang singularity
and primordial perturbations. In the new ekpyrotic scenario, the
bulk brane is absent and it is the hidden brane that collides
with the visible one, generating the hot Big Bang singularity.}
\label{fig:branes}
\end{figure}

In the ``old'' scenario~\cite{ekp}, the five dimensional bulk is also
assumed to contain various fields not described here, whose
excitations can lead to the spontaneous nucleation of yet another,
much lighter, freely moving, brane. In the so-called ``new''
scenario~\cite{ekpN}, and its cyclic extension~\cite{cyclic}, it is
the hidden boundary brane that is able to move in the bulk. In both
cases, this extra brane, if assumed BPS (as demanded by minimization
of the action) is flat, parallel to the boundary branes and initially
at rest. Non perturbative effects yield an interaction potential
between the visible and the bulk brane. The distance of the former to
the latter can be regarded as a scalar field living on the four
dimensional visible boundary brane whose effective action is thus that
of four dimensional GR together with a scalar field $\varphi$ evolving
in an exponential potential, namely
\begin{equation}\label{effec3D}
{\cal S}_{_4} = \int_{{\cal M}_4} \dd^4x \sqrt{-g_{_4}} \left[
{R_{_{(4)}}\over 2 \kappa} - {1\over 2}
\left(\partial\phi\right)^2 -V(\phi) \right],
\end{equation}
with
\begin{equation}
V(\varphi )=-V_{\rm i}\exp\biggr[-\frac{4\sqrt{\pi \gamma }}{
\mP}(\varphi -\varphi _{\rm i})\biggr],\label{expot}
\end{equation}
where $\gamma$ is a constant and $\kappa= 8\pi G =8\pi/\mP^2$. Apart
from the sign, the potential is the one that leads to the well known
power-law inflation model if the value of $\gamma$ lies in a given
range~\cite{powerlaw}.

The interaction between the two branes results in one (bulk or hidden)
brane moving towards the other (visible) boundary until they
collide. This impact time is then identified with the Big-Bang of
standard cosmology.  Slightly before that time, the exponential
potential abruptly goes to zero so the boundary brane is led to a
singular transition at which the kinetic energy of the bulk brane is
converted into radiation. The result is, from this point on, exactly
similar to standard big bang cosmology, with the difference that the
flatness problem is claimed to be solved by saying our Universe
originated as a BPS state (see however~\cite{cyclic}).

\begin{figure}[ht]
\includegraphics*[width=8.5cm, angle=0]{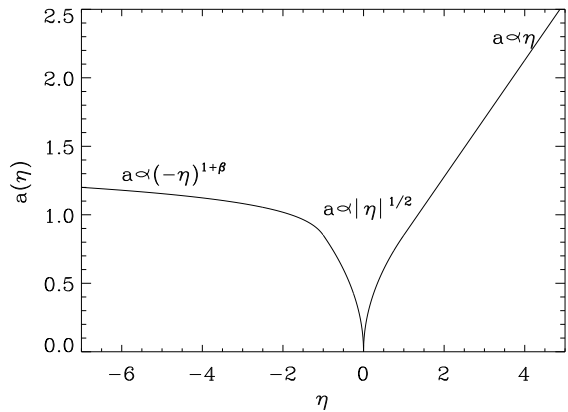}
\caption{Scale factor in the new ekpyrotic scenario. The Universe
starts its evolution with a slow contraction phase $a\propto
(-\eta)^{1+\beta}$ with $\beta=-0.9$ on the figure. The bounce
itself is explicitly associated with a singularity which is 
approached by the scalar field kinetic term domination phase, and the
expansion then connects to the standard Big-Bang radiation dominated
phase.}
\label{fig:anew}
\end{figure}

\section{Cosmological Perturbations in the New Ekpyrotic Model}
\label{sec:pert}

\subsection{The background}

As mentioned above, although the physics which describes the
evolution and the collision of the branes is very complicated, it
is assumed that it can be described by means of a simple
four-dimensional model. In this case, the equations that govern
the system are nothing but the Einstein equations
\begin{eqnarray}
\frac{3}{a^2}\biggl[\biggl(\frac{a'}{a}\biggr)^2 +{\cal
K}\biggr]&=&
\kappa \rho \label{G00}, \\ & & \nonumber \\
-\frac{1}{a^2}\biggl[2\biggl(\frac{a'}{a}\biggr)'
+\biggl(\frac{a'}{a}\biggr)^2+{\cal K}\biggr] &=& \kappa p,
\label{Gij}
\end{eqnarray}
where a prime denotes a derivative with respect to the conformal time
$\eta $. The Hubble parameter can be expressed as $H={\cal H}/a$ where
${\cal H}\equiv a'/a$. The equation of state $\omega$ can always be
written as
\begin{equation}
\omega \equiv \displaystyle{p\over\rho}=\frac{2\Gamma }{3}
\biggl(1+\frac{{\cal K}}{{\cal H}^2}\biggr)^{-1}-1,\label{eqstate}
\end{equation}
where the function $\Gamma (\eta )$ is defined by
\begin{equation}
\Gamma \equiv 1-\frac{{\cal H}'}{{\cal H}^2}+\frac{\Ka}{{\cal H}^2}.
\label{Gamma}\end{equation}
This last function is a direct generalization of the quantity $\gamma
\equiv 1-{\cal H}'/{\cal H}^2$ for non spatially flat universes since
it gives zero in the case of de Sitter spacetime. In the case of
spatially flat sections, the equation of state becomes $\omega
=2\gamma /3-1$. For a constant equation of state, the function
$\Gamma$ or $\gamma$ are constant. This is the case for the potential
(\ref{expot}) and the function $\gamma$ gets a constant value, 
explaining why we used the same symbol to denote these {\it
a priori} different objects. For the equation of state $\omega =-1$
(i.e., de Sitter space-time), they vanish.  Finally, the sound velocity
can be formally defined as
\begin{equation}\cS^2\equiv p'/\rho '.\label{cs2}\end{equation}
As already mentioned, in the ekpyrotic universe, it is
assumed that ${\cal K}=0$. As explained above, the pre-impact
phase consists in a scalar field dominated era and an
hydrodynamical era. We will follow in details the evolution of
the perturbations during these two eras.

\subsection{The scalar field era}

Let us start with the scalar field era. It is assumed that the
evolution of the four-dimensional background is governed by the
scalar field potential of Eq.~(\ref{expot}). This is a
well-studied case and the resolution of the Einstein equations
leads to a solution where the scale factor is a power-law of the
conformal time
\begin{eqnarray}
a(\eta ) &=& \ell _0(-\eta )^{1+\beta },
\\
\varphi (\eta ) &=& \varphi _{\rm i}+\frac{\mP }{2}
\sqrt{\frac{\gamma }{\pi }}(1+\beta )\ln (-\eta ).
\end{eqnarray}
As already mentioned, the function $\gamma $ is constant and its value
reads $\gamma = (2+\beta )/(1+\beta )$. In the ekpyrotic scenario, one
has $0< \beta +1\ll 1$. Since the Hubble parameter is given by
$aH=(1+\beta )/\eta <0$, this corresponds to a very slow contraction.
\par

In the case where a single scalar field dominates, the evolution
of density perturbations can be described by means of a single
equation,
\begin{equation}
\mu ''+\biggl[k^2-\frac{(a\sqrt{\gamma })''}{a\sqrt{\gamma }}
\biggr]\mu =0,
\end{equation}
where $k$ is the comoving dimensionless wavenumber and the quantity
$\mu $ is related to the Bardeen potential Fourier component $\Phi $
by the following relationship
\begin{equation}
\Phi = \frac{{\cal H}\gamma }{2k^2} \biggl(\frac{\mu
}{a\sqrt{\gamma }}\biggr)'.
\end{equation}
These quantities are related to the functions $v$ and $z$ in Eqs.~(19)
and (21) of Ref.~\cite{perturbekp} through $v\propto\mu$ and
$z\propto a\sqrt{\gamma}$. The initial condition for the function $\mu$
are fixed by the assumption that the quantum fluctuations are
initially placed in the vacuum state. This amounts to
\begin{equation}
\lim _{k/(aH)\rightarrow +\infty }\mu =-\frac{4\sqrt{\pi }}{\mP}
\frac{{\rm e}^{-ik(\eta -\eta _{\rm i})}}{\sqrt{2k}}.
\end{equation}
For a power-law scale factor, the equation of motion for the quantity
$\mu$ can be solved exactly in terms of Bessel functions as
$\mu=(k\eta)^{1/2} \left[ A_1(k) J_{\beta +1/2}(k\eta ) + A_2(k)
J_{-(\beta +1/2)}(k\eta )\right]$. In this case, the Bardeen potential
can be written as
\begin{widetext}
\begin{equation}
\Phi (\eta )=-\frac{{\cal H}\sqrt{\gamma }}{2ka}(k\eta )^{1/2}
\biggl[A_1(k)J_{\beta +3/2}(k\eta )- A_2(k)J_{-(\beta
+3/2)}(k\eta )\biggr],\label{Bessel}
\end{equation}
\end{widetext}
where the coefficients $A_1(k)$ and $A_2(k)$ are given by
\begin{eqnarray}
A_1(k) &=& \frac{\pi \sqrt{8}}{\mP \cos \beta \pi } \frac{{\rm
e}^{i(k\eta _{\rm i}-\pi \beta /2)}}{\sqrt{2k}},
\\
A_2(k) &=& -iA_1{\rm e}^{i\pi \beta }.
\end{eqnarray}
\par

Although these formulas are exact, there are not so easy to work
with and their interpretation is not especially illuminating. In
order to facilitate the interpretation, it is interesting to
proceed as follows. In the general case, i.e., even if the scale
factor does not behave as a power-law of the conformal time, the
quantity $\mu $ can be expressed as
\begin{equation}
\mu (\eta )=\sum _{n=0}^{\infty }b_{2n}(\eta )k^{2n},
\end{equation}
where the coefficients $b_{2n}(\eta )$ can be found by plugging
the previous equation in the equation of motion for $\mu $ and by
identifying the corresponding order in $k$. One finds
\begin{widetext}
\begin{eqnarray}
\mu (\eta )&=& \bar{A}_1a\sqrt{\gamma }\biggl[1-k^2\int ^{\eta }
\frac{{\rm d}\tau }{(a^2\gamma )(\tau )} \int ^{\tau }{\rm d}\tau
'(a^2\gamma )(\tau ') +{\cal O}(k^4)\biggr]
\nonumber \\
& &+\bar{A}_2a\sqrt{\gamma }\int ^{\eta }\frac{{\rm d}\tau
}{(a^2\gamma )(\tau)} \biggl[1-k^2\int ^{\tau }{\rm d}\tau
'(a^2\gamma )(\tau ') \int ^{\tau '}\frac{{\rm d}\tau
''}{(a^2\gamma )(\tau '')} +{\cal O}(k^4)\biggr].
\label{subtle}\end{eqnarray}
\end{widetext}
Inserting this expansion into the expression of $\Phi $, one
obtains at leading order
\begin{equation}
\Phi (\eta )=\frac{\bar{A}_2(k)}{2k^2}\frac{{\cal H}}{a^2}
-\frac{\bar{A}_1(k)}{2}\frac{{\cal H}}{a^2} \int ^{\eta }{\rm
d}\tau a^2\gamma .\label{modes}
\end{equation}
It is necessary to push the expansion to second order because we
see that the first term of $[\mu /(a\sqrt{\gamma })]'$ vanishes
in the expression of the Bardeen potential. Notice that Eq.~(22)
of Ref.~\cite{perturbekp} is not correct since one integration is
missing, see Eq.~(\ref{subtle}). The fact that it is necessary to
push the expansion up to second order to obtain the first
non-vanishing term in the Bardeen potential (the same is true for
the quantity $\zeta $, see below) has been called a subtlety in
Ref.~\cite{perturbekp} whereas this fact has been known for a long
time in the literature, see Ref.~\cite{perturb,MS}. One can easily
check that taking the limit $k\eta \rightarrow 0$ in
Eq.~(\ref{Bessel}) leads to the same dependence in conformal time
as in the previous equation. Explicitly, one has
\begin{equation}
\Phi (\eta )=-\frac{\bar{A}_2(k)}{2k^2}\frac{1+\beta }{\ell _0^2}
(-\eta )^{-3-2\beta } -\frac{\bar{A}_1(k)}{2}\frac{2+\beta
}{3+2\beta }.
\end{equation}
In an inflationary universe where $\beta \simeq -2$, the
$\bar{A}_1$-constant mode is the dominant mode since the
$\bar{A}_2$-mode decays as $-\eta$ when $\eta \rightarrow 0$. The
fact that it is proportional to $\beta +2$ and therefore vanishes
when $\beta =-2$ is just the well-known fact that there is no
density perturbations at all in a pure de Sitter phase. In the
ekpyrotic case, the situation is exactly the opposite, i.e., the
$\bar{A}_1$-constant mode is no longer the dominant mode. The
dominant mode is now the $\bar{A}_2$-mode since this one scales
like $-1/\eta$. The important point is the $k$-dependence of this
mode. This can be found by comparing the exact equation
(\ref{Bessel}) with Eq.~(\ref{modes}) which allows us to make the
link between the constants $\bar{A}_1$, $\bar{A}_2$ and $A_1$,
$A_2$. One obtains
\begin{equation}
\frac{\bar{A}_2(k)}{k^2}\sim k^{-\beta -5/2}, \quad
\bar{A}_1(k)\sim k^{\beta +1/2}.
\end{equation}
Therefore, for $\beta \simeq -1$, the dominant mode acquires a
scale invariant spectrum since $\Phi\propto k^{-3/2}$. Note
however in that respect that the ekpyrotic and de Sitter cases
are already at this stage very different since in the de Sitter
case, the scale invariant part of the spectrum is time
independent, contrary to what happens in the ekpyrotic situation.

\subsection{The hydrodynamical era}

\begin{figure}[t]
\includegraphics*[width=8.5cm, angle=0]{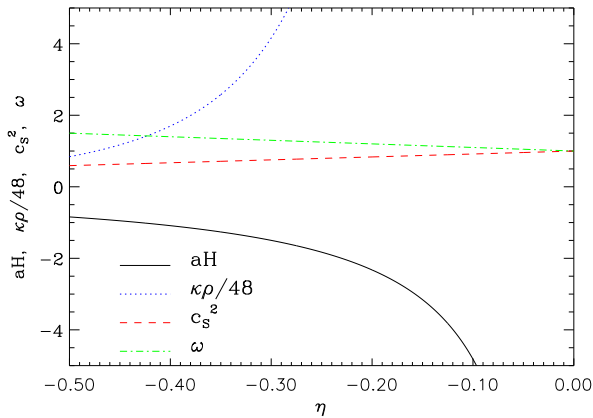}
\caption{Characteristic functions of the new ekpyrotic model [See
Eqs.~(\ref{wexp}), (\ref{Hekp}), (\ref{rhoekp}) and
(\ref{cs2ekp})]. The parameters are chosen as $\ell _0=1$ and $\omega
_1=-1$ (the ekpyrotic scenario requires $\omega_1$ 
to be negative).} \label{fig:exact}
\end{figure}

As the bulk brane is approaching the visible brane, the scalar field
blows up. In the ekpyrotic scenario, it is assumed that the shape of
the potential changes and goes to zero as the collision is taking
place. Consequently, just before the collision, the equation of state
tends toward a stiff equation of state, i.e., $\omega_0 =1$. More
generally, Ref.~\cite{perturbekp} considers the situation where
\begin{equation}
\omega \equiv {p\over \rho} =\omega_0 + \omega _1\eta +\omega _2\eta
^2 +\cdots . \label{wexp}
\end{equation}
Being given a general equation of state $\omega (\eta )$, it is
easy to find the corresponding scale factor. It reads
\begin{equation}
a(\eta )=\ell _0\exp\biggl(\int ^{\eta }{\rm d}\tau \biggl\{C+
\frac{1}{2}\int ^{\tau }{\rm d}\tau '[1+3\omega (\tau ')]
\biggr\}^{-1}\biggr),
\end{equation}
where $\ell _0$ and $C$ are arbitrary constants. In this article,
without any essential loss of generality, we restrict our
considerations to the truncated equation of state $\omega =1 +\omega
_1\eta$, i.e., $\omega_0=1$. In this case the scale factor can be
integrated explicitly using Eq.~(2.154) of Ref.~\cite{Grad} and the
result reads
\begin{equation}
a(\eta )=\ell _0\left| \frac{\eta }{2(1+3\omega_0)+3\omega _1\eta
}\right|^{2/(1+3\omega_0)}.
\end{equation}
To our knowledge, this is a new solution. Moreover, as we will show,
all the relevant perturbed quantities can be calculated exactly in
this model. Therefore, this also constitutes a new exactly integrable
case for the theory of cosmological perturbations. In the previous
formula the constant $C$ has been chosen such that $a(0)=0$ in
agreement with the new ekpyrotic scenario. There is a divergence when
$\omega _1\eta =-8/3$. This just signals that more terms should be
included in the Taylor series of the equation of state which anyway,
under the form of Eq.~(\ref{wexp}), is only valid for in the vicinity
of the bounce.  Moreover, in the collapsing phase, one has
$\omega_1<0$ with $\eta<0$ so the model is divergence free in the pre
impact era. It is interesting to see how the physical quantities that
characterize the model behave. We find
\begin{eqnarray}
{\cal H}(\eta ) &=& \frac{4}{\eta (8+3\omega _1\eta )},\label{Hekp} \\
\kappa \rho (\eta ) &=& \frac{48}{\ell _0^2|\eta ^3(8+3\omega _1\eta )|} ,
\label{rhoekp} 
\\
\cS^2 (\eta ) &=& \frac{5}{6}+\frac{3}{4}\omega _1\eta +
\frac{1}{6+3\omega _1\eta },\label{cs2ekp}
\end{eqnarray}

\noindent where the ``sound velocity'' is given by
Eq.~(\ref{cs2}). As expected, the Hubble parameter and the energy
density blow up at $\eta =0$. These functions are shown in
Fig.~\ref{fig:exact}
On the contrary, the equation of state and the sound velocity are
regular. The fact that the sound velocity is $1$ at $\eta =0$ can
easily be understood. This is a consequence of the equation
\begin{equation}
\omega '=-3{\cal H}(\cS^2-\omega )(1+\omega ).
\end{equation}
When $\eta\to 0$ and $\omega\not= -1$, it is necessary that $\cS^2$
behaves as $\cS^2\to\omega$ in order to obtain a finite $\omega'$ as
$\Hu$ diverges in this limit.

\begin{figure}[ht]
\includegraphics*[width=8.5cm, angle=0]{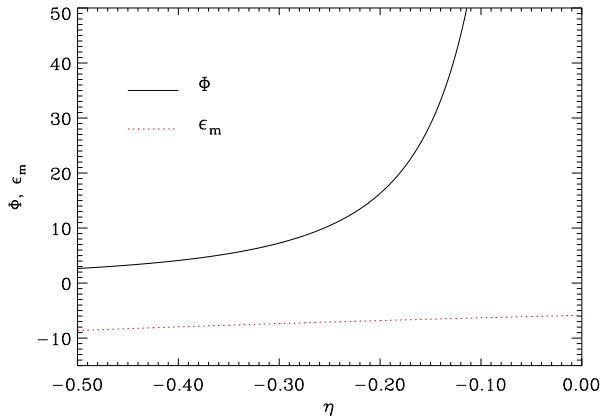}
\caption{The divergence in the Bardeen potential and the gauge
invariant energy density perturbation near the bounce in the new
ekpyrotic scenario. Parameters are chosen as $\ell
_0=1$ and $\omega _1=-1$. The Bardeen potential $\Phi$ is from
Eq.~(\ref{Phidiv}) with $s B_1=B_2=1$, while $k^2s B_1=k^2B_2=1$ for
$\epsilon_{_{\rm m}}$ in Eq.~(\ref{solepsm}).} \label{fig:eps}
\end{figure}

In the phase dominated by an (effective) hydrodynamical fluid,
the equation that governs the evolution of density perturbations
reads
\begin{widetext}
\begin{equation}
\Phi '' +3(1+\cS^2){\cal H}\Phi' +[2{\cal H}'+(1+3\cS^2) ({\cal
H}^2-\Ka)]\Phi +\cS^2(k^2-\Ka)\Phi=0,\label{Bardeen}
\end{equation}
where we have assumed that there is no entropy production. This
equation can also be put under the form of an equation of motion
for a parametric oscillator. Let us define $\mu$ and $\theta$ by
\begin{equation}
\mu \equiv \frac{2a^2\theta }{3{\cal H}}\Phi, \quad \theta \equiv
\frac{1}{a}\biggl(\frac{\rho }{\rho +p}\biggr)^{1/2}
\biggl(1-\frac{3\Ka}{\kappa \rho a^2}\biggr)^{1/2}=\frac{1}{a}
\sqrt{\frac{3}{2\Gamma }},\label{mutheta}
\end{equation}
where $\Gamma$ is defined through Eq.~(\ref{Gamma}). Of course, in the
ekpyrotic case, these equations should be used with ${\cal K}=0$,
although we wrote them here in their full generality since we will use
them in the regular $\Ka =1$ case in the following section. Then, the
equation of motion of $\mu$ can be written under the form of a
parametric oscillator equation of motion
\begin{equation}
\mu ''+\biggl[\cS^2(k^2-\Ka)-\frac{\theta ''}{\theta }\biggr]\mu
=0.\label{pot}
\end{equation}
As for the scalar field case, one can solve this equation
perturbatively. The solution now reads
\begin{eqnarray}
\mu (\eta ) &=& B_1(k)\theta \biggl[1-k^2\int ^{\eta } \frac{{\rm
d}\tau }{\theta ^2(\tau )} \int ^{\tau }{\rm d}\tau '(\cS^2\theta
^2)(\tau ') +{\cal O}(k^4)\biggr]
\nonumber \\
& &+B_2(k)\theta \int ^{\eta }\frac{{\rm d}\tau }{\theta ^2}
\biggl[1-k^2\int ^{\tau }{\rm d}\tau '(\cS^2\theta ^2)(\tau ')
\int ^{\tau '}\frac{{\rm d}\tau ''}{\theta ^2(\tau '')} +{\cal
O}(k^4)\biggr].\label{musol}
\end{eqnarray}
\end{widetext}
In the long-wavelength approximation, the solution is
\begin{equation}
\Phi =\frac{3}{2}B_1(k)\frac{{\cal H}}{a^2}
+\frac{3}{2}B_2(k)\frac{{\cal H}}{a^2} \int ^{\eta }\frac{{\rm
d}\tau }{\theta ^2}.\label{Phisol}
\end{equation}
As expected, at this order, the solution does not depend on the
sound velocity. In the above equation, the integral can be easily
performed. The Bardeen potential reads
\begin{widetext}
\begin{equation}
\Phi (\eta )= \frac{6s B_1(k)}{\ell _0^2\eta ^2}
+\frac{B_2(k)}{9 (\omega _1\eta)^2}[-12 \omega _1\eta +9(\omega
_1\eta )^2+32\ln (8+3\omega _1\eta )], \label{Phidiv}
\end{equation}
\end{widetext}
where $s$ is the sign of the conformal time $\eta$. The Bardeen
potential blows up as $\eta $ is approaching zero and the linear
theory becomes meaningless. In Ref.~\cite{perturbekp}, it is argued
that $\Phi (\eta ) $ should not be used. Instead, it is proposed to
use the density contrast~\cite{Bardeen} $\epsilon _{_{\rm m}}\equiv
\delta \rho /\rho $ which is linked to the Bardeen potential by the
relation
\begin{equation}
\epsilon _{_{\rm m}}=\frac{k^2\Phi }{\rho a^2}.
\end{equation}
Then the superhorizon solution of $\epsilon _{\rm m}$ can be expressed as
\begin{equation}
\epsilon _{_{\rm m}}=\frac{k^2B_1(k)}{2{\cal H}a^2}
+\frac{k^2B_2(k)}{2{\cal H}a^2} \int ^{\eta }\frac{{\rm d}\tau
}{\theta ^2}.
\end{equation}
Explicitly, the solution can be written as
\begin{widetext}
\begin{equation}
\epsilon _{_{\rm m}}=\frac{8k^2s B_1(k)}{\ell _0^2}\left[1+
\frac{3}{4}\omega _1\eta +\frac{9}{64}(\omega _1\eta )^2\right]
+\frac{4k^2B_2(k)}{27\omega _1^2} \left[-12 \omega _1\eta +9(\omega
_1\eta )^2+32\ln (8+3\omega _1\eta )\right] \left[1+\frac{3}{4}\omega
_1\eta +\frac{9}{64}(\omega _1\eta )^2\right] ,\label{solepsm}
\end{equation}
in which the limit $\omega_1\to 0$, being singular, is not applicable.
We see that the variable $\epsilon _{_{\rm m}}$ is regular at $\eta
=0$ because the divergence has been canceled by the factor $1/(\rho
a^2)$. Note however that for a constant equation of state
$\omega=\omega_0$, this variable is not regular if $-1/3 \leq \omega_0
< 1$. The regularity of $\epsilon _{_{\rm m}}$ thus depends on the
matter content as the singularity is approached.  If we expand the
above equation around $\eta =0$, we find
\begin{equation}
\epsilon _{_{\rm m}}=\biggl[\frac{8k^2s B_1}{\ell _0^2}
+\frac{128k^2B_2}{9\omega _1^2}\ln
2\biggr]\biggl[1+\frac{3}{4}\omega_1\eta +{\cal O}(\eta
^3)\biggr]+\biggl[\frac{9k^2s B_1\omega _1^2}{8\ell _0^2}
+k^2B_2\left(1+2\ln 2\right)\biggr]\biggr[\eta ^2 +{\cal O}(\eta
^3)\biggr].\label{solepsmexp}
\end{equation}
\end{widetext}
This equation is in agreement with Eq.~(42) of Ref.~\cite{perturbekp}
with $\omega _2=0$. For $k\not= 0$, the first ${\cal O}(\eta^3)$ is
replaced with ${\cal O}(k^2 \eta^2 \ln |\eta|)$.  The Bardeen potential
and the density contrast are plotted in Fig.~\ref{fig:eps}.
In Ref.~\cite{perturbekp}, the solution $\epsilon _{_{\rm m}}$ has
not been expanded in the basis of the growing and decaying mode
but has been written as
\begin{equation}
\epsilon _{_{\rm m}}=\epsilon _0(k)D(k,\eta ) +\epsilon
_2(k)E(k,\eta ),
\end{equation}
where $D\equiv 1+3\omega _1\eta /4 +{\cal O}(k^2 \eta^2 \ln |\eta|)$
and $E\equiv \eta ^2+{\cal O}(\eta ^3)$. The link between the
coefficients of the growing and decaying modes $B_1$, $B_2$ and the
coefficients $\epsilon _0$ and $\epsilon _2$ of the $(D,E)$ basis is
obvious
\begin{eqnarray}
\epsilon _0(k) &=& \frac{8k^2s B_1}{\ell _0^2}
+\frac{128k^2B_2}{9\omega _1^2}\ln 2, \label{E0}
\\
\epsilon _2(k) &=& \frac{9k^2s B_1\omega _1^2}{8\ell _0^2}
+k^2B_2\left(1+2\ln 2\right).\label{E2}
\end{eqnarray}
At leading order in $k$, the previous equations imply $\epsilon
_2=3\epsilon _0 w^{(2)}/8$, in agreement with the equation in the last
line of the paragraph below Eq.~(45) of Ref.~\cite{perturbekp}, being
given that in the present context the variable $w^{(2)}$ of
Ref.~\cite{perturbekp} is simply $3\omega _1^2/8$; this means that
$\epsilon _0$ and $\epsilon _2$ are of the same order in $k$. The
inverse transformation reads
\begin{eqnarray}
s B_1(k) &=& \frac{\ell _0^2}{8k^2}\left(1+2\ln 2\right)\epsilon _0
-\frac{16\ell _0^2\ln 2}{9\omega _1^2k^2}\epsilon _2, \label{B1}
\\
B_2(k) &=& \frac{1}{k^2}\biggl(-\frac{9\omega _1^2}{64}\epsilon
_0+ \epsilon _2\biggr).\label{B2}
\end{eqnarray}
Let us notice that if we want to obtain the other terms of the
expansion, we need to use the higher order terms in the
expression~(\ref{musol}) of $\mu (\eta )$. For example the first
(respectively second) branch next-to-leading order term can be
expressed in terms of elementary functions and of dilogarithm Li$_2
(\eta)$ [resp. trilogarithm Li$_3 (\eta)$] functions~\cite{dilog}. An
expansion of these functions around $\eta=0$ reproduces Eq.~(43) of
Ref.~\cite{perturbekp}.

Finally, let us end this section by a discussion on the quantity
$\zeta (\eta )$. This one is defined by the following equation
\begin{eqnarray}
\zeta & \equiv & \frac{2}{3}\frac{{\cal H}^{-1}\Phi
'+\Phi}{1+\omega }+\Phi\label{def:zeta}
\\
&=& \frac{{\cal H}^2+\Ka}{{\cal H}^2}\theta ^2\biggl(\frac{\mu
}{\theta }\biggr)' +\frac{3{\cal H}\mu }{2a^2\theta }
\biggl[1-\frac{\gamma }{\Gamma }\biggl(1+\frac{{\cal K}}{{\cal
H}^2}\biggl) \biggr].\ \label{zetamu}
\end{eqnarray}
Let us briefly recall under which conditions, the quantity $\zeta $
can be considered as a constant. First of all, ${\cal K}$ should be
equal to zero such that the last term in the above equation
disappears. Secondly, there should be no entropy
perturbations. Thirdly, only the growing mode should be considered and
we note that it is crucial to discard the singular mode~\cite{MS}. In
order to obtain the explicit expression of $\zeta$, one can insert the
formula (\ref{musol}) giving $\mu $ in Eq.~(\ref{zetamu}). As is well
known~\cite{MS}, it is necessary to push the expansion to second order
because we see that the first term of $(\mu /\theta )'$ vanishes. One
obtains
\begin{equation}
\zeta =-B_1k^2\int ^{\eta }{\rm d}\tau \cS^2(\tau )\theta
^2(\tau ) +B_2 +{\cal O}(k^2).\label{zetasol}
\end{equation}
For the exact model studied here the integral in the above
equation can be performed exactly. The result reads
\begin{equation}
\zeta =-\frac{k^2s B_1}{12\ell _0^2}\biggl[27\omega _1\eta
+\frac{4}{2+\omega _1\eta }+48\ln (-\eta )\biggr]+B_2.
\end{equation}
Therefore, this quantity has a logarithmic divergence as the
point where the scale factor vanishes is approached. This is in
agreement with the analysis of Ref.~\cite{perturbekp}. The divergence 
is again a signal that the linear theory looses any meaning.

\subsection{Matching conditions}

We have at our disposal the solutions in each era. The goal is now to
join them. The first step is to pass from the scalar field era to the
hydrodynamical era. Since the equation of state can be made continuous
at this point, one has $[a]=[a']=[a'']=0$. In this case, the usual
joining conditions can be used and the Bardeen potential and its
derivative are continuous. This means that the growing mode in the
hydrodynamical era acquires a scale invariant spectrum. In other
words, $B_1(k)\sim k^{-3/2}$ and $B_2\sim k^{-1/2}$ because $\Phi$ has
the same shape in both eras. The same applies to the other transition,
in the expanding regime, from domination by the scalar field kinetic
term to the radiation epoch.  \par

Clearly, the non trivial step is how to propagate the perturbations
through the singularity. We have to connect the solution in the
pre-impact hydrodynamical phase with equation of state $\omega
=1+\omega _1^<\eta $ with the solution in the post-impact phase with
$\omega =1+\omega _1^>\eta $, being given that $\omega _1^<\neq \omega
_1^>$. {\it A priori}, this seems simply impossible because the theory
({\it a fortiori} the linear theory) looses any meaning (signaled by
the divergence of the scalar curvature and/or of the Bardeen
potential): how to perturb around a singular background? Even if we
are ready to accept this, the theory suffers from a serious
trans-Planckian problem since all the wavelengths become at some point
smaller than the Planck length~\cite{transPl}. However, despite these 
seemingly
insurmountable difficulties, Ref.~\cite{perturbekp} goes on along the
following lines. The fact that the quantity $\epsilon _{_{\rm m}}$ is
regular is used in an essential way. A first approach would be to
impose $[\epsilon _{_{\rm m}}]=[\epsilon _{_{\rm m}}']=0$. The first
condition means $\epsilon _0^>=\epsilon _0^<$ whereas the second
cannot be applied since $\epsilon _{_{\rm m}}'(0)= 3\omega _1/4$,
which is required to be different in the pre- and post impact
eras since $\omega _1^<\neq \omega _1^>$. The fact that the 
derivative cannot be made continuous (contrary
to the claims of Ref.~\cite{perturbekp}) can be directly traced back
to the fact that the background is singular. As a consequence, one
cannot find $\epsilon _2^>$.  Then, a new suggestion is given to find
the coefficients $\epsilon _0^>$ and $\epsilon _2^>$. It consists in
assuming that
\begin{equation}
\epsilon _0^>(k)=\epsilon _0^<(k), \qquad \epsilon
_2^>(k)=\epsilon _2^<(k) ,\label{conteps}
\end{equation}
i.e., one assumes that the energy density perturbation as well as the
second derivative of $(\epsilon_{_{\rm m}}-\epsilon_0 D)$ are
continuous across the bounce. At this point, we would like to stress
the following remark: the usual matching conditions stem from a well
defined geometrical requirement~\cite{MS,matching}. In physics, in
general, the requirement is made that the function and its derivative
should be continuous because the point considered is almost never
considered to be a singular point. The situation here is therefore
extremely special as there does not appear to be any physical reason
to enforce any matching conditions, especially on a variable which,
although finite, is the perturbation of a diverging background
quantity. Let us however press on to assert if they lead, as claimed
in Ref.~\cite{perturbekp}, to a unique spectrum: this fact in itself
would maybe justify {\it a posteriori} this choice for the criterion.

Of course, in the post-impact phase, one is interested in the
coefficient of the growing mode, i.e., in the spectrum of $\Phi$
perturbations, and not in the coefficients $\epsilon _0^>$ and
$\epsilon _2^>$. This is because the growing mode directly provides us
with the spectrum. Plugging Eqs.~(\ref{E0}) and (\ref{E2}) into
Eq.~(\ref{B2}), and using the continuity condition~(\ref{conteps})
permits to evaluate this spectrum. The result reads
\begin{widetext}
\begin{eqnarray}
B_1^>(k)&=&-B_1^{<} (k)\left[1+2\left(1-\frac{\omega _1^{<2}}{\omega
_1^{>2}} \right)\ln 2\right]+\frac{16\ell _0^2\ln 2}{9} B_2^<(k)
(1+2\ln 2)\left( {1\over \omega _1^{<2}}-{1\over \omega
_1^{>2}}\right).\\
B_2^>(k)&=&-\frac{9}{8\ell _0^2}B_1^<(k)(\omega _1^{<2}-\omega
_1^{>2}) +B_2^{<}(k)\left[1+2\left(1-\frac{\omega _1^{>2}}{\omega
_1^{<2}} \right)\ln 2\right].
\end{eqnarray}
\end{widetext}
Therefore, in the limit of long wavelengths, the dominant term is
$B_1^<(k)$ and we have $B_2^>(k)\sim k^{-3/2}$ at least as long as
$\omega _1^{<2}\neq \omega _1^{>2}$. In this case, the spectrum of the
Bardeen potential is scale-invariant in the post-impact phase. Let us
now study the new prescription in greater details. It is clear that
the function $\epsilon _{_{\rm m}}$ can also be written as
\begin{eqnarray}
\epsilon _{_{\rm m}} &=& \epsilon _0(k)D(k,\eta ) +\frac{\epsilon
_2(k)}{f(\omega _1,k)}f(\omega _1,k) E(k,\eta ),
\\
&=& \bar{\epsilon }_0(k)\bar{D}(k,\eta ) +\bar{\epsilon
}_2(k)\bar{E}(k,\eta ),
\end{eqnarray}
where $\bar{\epsilon }_0(k)=\epsilon _0(k)$, $\bar{D}(k,\eta )=
D(k,\eta )$, $\bar{\epsilon }_2(k)=\epsilon _2(k)/f(\omega
_1,k)$ and $\bar{E}(k,\eta )=f(\omega _1,k)E(k,\eta )$. Of course
the choice of the basis has no physical meaning at all and we can
equally well expand $\epsilon _{_{\rm m}}(\eta )$ in the basis
$(D,E)$ or $(\bar{D},\bar{E})$. Let us remark that we could
choose a more general change of basis but in the present context
the continuity of $\epsilon _{_{\rm m}}$ would no longer be
guaranteed. In the standard case, such a change of basis has
obviously no consequence on the final spectrum as it should. As
we are going to show, this is not the case for the new proposal of
Ref.~\cite{perturbekp}. With the new basis, the matching
conditions of Eq.~(\ref{conteps}) transforms into
\begin{equation}
\bar{\epsilon }_0^>(k)=\bar{\epsilon }_0^<(k), \qquad
\bar{\epsilon }_2^>(k)=\bar{\epsilon }_2^<(k) .
\end{equation}
This leads to the following expression for the coefficient
$B_2^>(k)$
\begin{widetext}
\begin{equation}
B_2^>(k)=-\frac{9}{8\ell _0^2}B_1^<(k)\left(\frac{f^>}{f^<} \omega
_1^{<2}-\omega _1^{>2}\right) +B_2^{<}(k)\left[\frac{f^>}{f^<}
+2\left(\frac{f^>}{f^<} -\frac{\omega _1^{>2}}{\omega _1^{<2}}
\right)\ln 2\right].
\end{equation}
\end{widetext}
Now since the choice of $f$ is completely arbitrary, one can always
choose $f^>=\omega _1^{>2}$ and $f^<=\omega _1^{<2}$. In this case the
first term cancels out and we are left with a spectrum $B_2^>(k)\sim
k^{-1/2}$ corresponding to a spectral index $\nS=3$. More
generally, one is free to choose the function $f$ as a function of $k$
and in this case the spectrum is also completely changed.  \par In
conclusion, the proposal of Ref.~\cite{perturbekp} rests on a
non-physical choice. Choosing the normalization of the function $E$
such that it leads to a scale invariant spectrum seems to be
arbitrary. The standard junction conditions do not depend on the
normalization of the mode functions.  So even if one admits that we
can somehow pass through a singularity, it seems that there is no
convincing way to find a scale-invariant spectrum. This is probably
because through a singularity any result can be obtained.

\subsection{Testing the matching conditions: the radiation to matter
transition}

In a recent proposal~\cite{Ruth}, it was argued that the junction
conditions advocated in the previous section could be expressed in a
very similar way to the usual junction conditions. It is
well-known~\cite{mtw} that matching conditions follow from the
requirement that $[h_{ij}]=0$ and $[K_{ij}]=0$ where $h_{ij}$ is the
metric of the spacelike sections and $K_{ij}$ is the associated second
fundamental form. The question is then: on which surface should these
conditions be imposed? The standard answer is to match on a surface of
constant longitudinal gauge (gauge-invariant) energy density denoted
$\epsilon_{\rm g}$ by Bardeen~\cite{Bardeen}. The proposal of
Ref.~\cite{Ruth} is to perform the matching on a surface of constant
comoving (gauge-invariant) energy density denoted $\epsilon_{\rm
m}$. This is the quantity used above and advocated
in~\cite{perturbekp}.

The standard junction conditions reduce to the continuity of the
Bardeen potential and $\zeta$, $[\Phi]=[\zeta]=0$. The new ones amount
to $[\Phi]=[\Hu \Phi + \Phi']=0$; see Eqs.~(20) and (21) of
Ref.~\cite{Ruth}. We have taken the surface layer pressure to be zero
as it was argued in Ref.~\cite{Ruth} that this does not play a crucial
role in the present context. However, we will come back to this point
shortly. If $[a]=[a']=[a'']=0$, the two sets of conditions are
equivalent and both lead to $[\Phi]=[\Phi']=0$ as already
mentioned. On the other hand, there exists a situation for which the
two sets are not equivalent, namely that of a sharp transition, i.e.,
one for which the equation of state $\omega$ jumps.  In order to
discuss the accuracy of the new proposal, let us examine the case of
the radiation to matter transition. The advantage is that the exact
solution is known and then we can compare whether the different set of
junction conditions reproduce or not the correct result.

In the radiation to matter domination transition, Einstein equations
can be solved exactly and the scale factor is given by the following
expression~\cite{perturb}
\begin{equation}
a(\eta )=a_{\rm eq}\biggl[b^2
\biggl(\frac{\eta }{\eta _{\rm eq}}\biggr)^2
+2b\biggl(\frac{\eta }{\eta _{\rm eq}}\biggr)\biggr].
\label{aradmat}\end{equation}
For $\eta \ll \eta _{\rm eq}$, the scale factor is approximatively
linear in the conformal time and the universe is radiation dominated
whereas for $\eta \gg \eta _{\rm eq}$ it is quadratic in the conformal
time and the universe is matter dominated. The freely adjustable
coefficient $b=\sqrt{2}-1$ is chosen such that $a(\eta =\eta _{\rm
eq}) =a_{\rm eq}$ (note that the different choice $b=1$ is made
in~\cite{perturb}). The superhorizon solution for the Bardeen potential is
Eq.~(\ref{Phisol}) which, in the case of the radiation-matter
transition~(\ref{aradmat}), can be written as
\begin{widetext}
\begin{equation}
\Phi (\eta ) = \biggl(\frac{3b}{\eta _{\rm eq}a_{\rm
eq}^2}B_1\biggr)\frac{b\eta /\eta _{\rm
eq}+1}{b^3 (\eta /\eta _{\rm eq})^3(b\eta /\eta
_{\rm eq}+2)^3} + B_2\frac{b\eta /\eta
_{\rm eq}+1} {(b\eta /\eta _{\rm eq}+2)^3}
\biggl[\frac{3}{5}b^2\biggl(\frac{\eta }{\eta _{\rm
eq}}\biggr)^2 +3b\biggl(\frac{\eta }{\eta _{\rm eq}}\biggr)
+\frac{13}{3} +\frac{1}{b\eta/\eta _{\rm eq}+1}\biggr],
\label{potradmat}\end{equation}
\end{widetext}
in which the lower bound of the integral in Eq.~(\ref{Phisol}) has
been chosen to cancel the $\Hu/a^2$ contribution of the second
branch. From this expression, it is easy to check that, for the
growing mode, one has
\begin{equation}
\frac{\Phi (\eta \gg \eta _{\rm eq})}{\Phi (\eta \ll \eta _{\rm
eq})} =\frac{9}{10}. \label{ratio}
\end{equation}
This result is nothing but the standard result of the inflationary
cosmology, applied to the radiation to matter transition. In the same
manner, the quantity $\zeta $ can be calculated exactly. One obtains
\begin{equation}
\zeta (\eta )=B_2.
\end{equation}
This result is valid as soon as the decaying mode, not taken into
account here, had enough time to decay.

Let us now turn to the piecewise solution for which the same situation
can also be described by means of the following approximation for the
scale factor
\begin{eqnarray}
a^<(\eta ) &=& \frac{a_{\rm eq}}{\eta _{\rm eq}}\eta , 
\\
a^>(\eta ) &=& \frac{a_{\rm eq}}{4}
\biggl(1+\frac{\eta }{\eta _{\rm eq}}\biggr)^2,
\end{eqnarray}
where $[a]=[a']=0$ has been imposed in agreement with the background
junction conditions. For each region, the exact superhorizon solution
for the Bardeen potential can easily be obtained and reads
\begin{eqnarray}
\Phi ^<(\eta ) &=& \frac{3B_1^<}{2\eta _{\rm eq}a_{\rm eq}^2}
\biggl(\frac{\eta }{\eta _{\rm eq}}\biggr)^{-3}+\frac{2}{3}B_2^<,
\\
\Phi ^>(\eta ) &=& \frac{48B_1^>}{\eta _{\rm eq}a_{\rm eq}^2}
\biggl(1+\frac{\eta }{\eta _{\rm eq}}\biggr)^{-5}+\frac{3}{5}B_2^>.
\end{eqnarray}
Similarly, one gets the quantity $\zeta$ as
\begin{equation}
\zeta ^<(\eta )=B_2^<, \quad \zeta ^>(\eta )=B_2^>,
\end{equation}
where, as emphasized above, the decaying mode is assumed negligible.

\begin{figure}[ht]
\includegraphics*[width=8.5cm, angle=0]{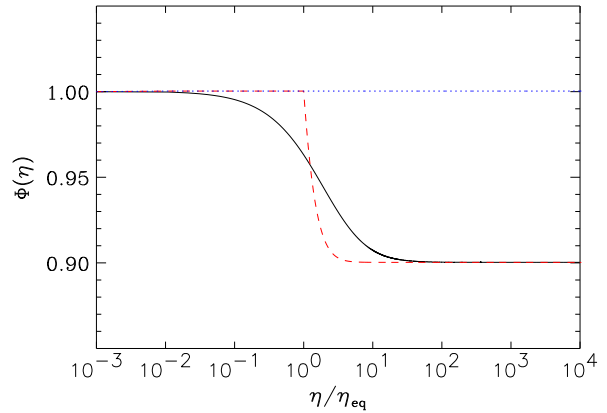}
\caption{Time evolution of the Bardeen potential during the radiation
to matter domination transition. The full line shows the exact
solution~(\ref{potradmat}), the dashed line represents the usual
approximation~(\ref{PhiUsual}), and the dotted line is obtained
using the new proposal~(\ref{PhiRuth}) for the junction
conditions. This last approximation is not in fact constant but, as
discussed below Eq.~(\ref{PhiRuth}), only hardly varying at all on
the scale shown. In order to ensure that initial conditions are
identical for the three curves, we have set numerically $\Phi_{\rm i}
= \Phi_{\rm i}' = 1$ [see above Eq.~(\ref{phii})], and used $B_1
\simeq -b^2 a^2_{\rm eq} \eta_{\rm i}^4 (b \Phi_{\rm i} + 8 \eta_{\rm
eq} \Phi_{\rm i}')/9\eta_{\rm eq}^2$ and $B_2\simeq 3/2 (\Phi_{\rm i}
+ \eta_{\rm i} \Phi_{\rm i}'/3)$ in Eq.~(\ref{potradmat}) valid in the
limit $\eta_{\rm i}\ll \eta_{\rm eq}$.}
\label{fig:radmat}
\end{figure}

We are now in the position to relate the various quantities of
interest before and after the transition. For this purpose, let us now
apply to set of junction conditions. The standard matching conditions
stipulate that $[\Phi ]=[\zeta ]=0$. This amounts to
\begin{eqnarray}
B_1^> &=& B_1^< +\frac{2\eta _{\rm eq}a_{\rm eq}^2}{45}B_2^< ,
\\
B_2^> &=& B_2^<.
\end{eqnarray}
For a sharp transition having $[\Hu]=0$ and $[a'']\not=0$, implying
$[p]\not=0$, the matching conditions proposed in Ref.~\cite{Ruth} are
equivalent to $[\Phi ]=[\Phi ']=0$. From the very
definition~(\ref{def:zeta}) of $\zeta$, these conditions, together
with $[\omega]\not=0$, implies $[\zeta]\not=0$. Therefore, as
announced, the two sets of junction conditions are not equivalent.
Applying the new matching procedure yields
\begin{eqnarray}
B_1^> &=& \frac{6}{5}B_1^< , \\
B_2^> &=& -\frac{1}{2\eta _{\rm eq}a_{\rm eq}^2} B_1^<
+\frac{10}{9}B_2^<.
\end{eqnarray}
In turn, the coefficients $B_1^<$ and $B_2^<$ are fixed by the initial
conditions at some time $\eta _{\rm i}$. Let us express these
coefficients in terms of $\Phi _{\rm i} $ and $\Phi '_{\rm i}$, the
Bardeen potential and its derivative at some initial time $\eta_{\rm
i} \ll \eta_{\rm eq}$ respectively. Before the transition, the result
is
\begin{equation}
\Phi ^<(\eta )=\Phi _{\rm i}+\frac{1}{3}\eta _{\rm i}\Phi _{\rm i}'
-\frac{1}{3}\eta _{\rm i}\Phi _{\rm i}'
\frac{\eta _{\rm i}^3}{\eta ^3}.
\label{phii}\end{equation} 
If one uses the standard junction conditions, the Bardeen potential
after the transition can be written as
\begin{eqnarray}
\Phi ^>(\eta ) &=& \frac{16}{5}\biggl(\Phi _{\rm i}+
\frac{1}{3}\eta _{\rm i}\Phi _{\rm i}'
-\frac{10}{3}\eta _{\rm i}\Phi _{\rm i}'
\frac{\eta _{\rm i}^3}{\eta _{\rm eq}^3}\biggr)
\biggl(1+\frac{\eta }{\eta _{\rm eq}}\biggr)^{-5}
\nonumber \\
& & +\frac{9}{10}\biggl(\Phi _{\rm i}+
\frac{1}{3}\eta _{\rm i}\Phi _{\rm i}'\biggr),\label{PhiUsual}
\end{eqnarray}
leading to the correct ratio as in Eq.~(\ref{ratio}). On the other
hand, the matching conditions proposed in Ref.~\cite{Ruth} lead to
\begin{eqnarray}
\Phi ^>(\eta ) &=& -\frac{64}{5}\eta _{\rm i}\Phi _{\rm i}'
\frac{\eta _{\rm i}^3}{\eta _{\rm eq}^3}
\biggl(1+\frac{\eta }{\eta _{\rm eq}}\biggr)^{-5}
+\Phi _{\rm i}
\nonumber \\
& & +\frac{1}{3}\eta _{\rm i}\Phi _{\rm i}'
\biggl(1+\frac{1}{5}\frac{\eta _{\rm i}^3}{\eta _{\rm eq}^3}
\biggr).\label{PhiRuth}
\end{eqnarray}
The ratio between the constant parts of the Bardeen potential before
and after the transition is then very close to unity, namely $\simeq 1
+ \eta_{\rm i}\Phi _{\rm i}'/(3\Phi_{\rm i})$. Fig.~\ref{fig:radmat}
illustrates this point.

The new proposal does not reproduce the exact result for the radiation
to matter domination transition. One could argue however that the
situation for which the new junction conditions were suggested is
different from the standard case, and that therefore new rules must be
applied. This would require different physical prescriptions for
different situations, whereas it seems to us that a unified approach is
more satisfactory.

\section{Hydrodynamical bounce and the conserved quantity $\zeta $}
\label{sec:hydro}

We now turn to the second topic of this work in relation with the
bouncing phase. From now on, we shall consider a regular bounce, i.e.,
one such that the scale factor never vanishes. As already mentioned,
it is clear that such a bounce cannot be described in the same
framework as in the ekpyrotic case because it is impossible to have a
bounce if ${\cal K}=0$ within GR. For instance, this can be seen if
matter consists of a single scalar field since Einstein equations yield
\begin{equation} \rho+p = {2\over
\kappa a^2} (\Hu^2 - \Hu'),
\end{equation}
which shows that $\rho+p$ should be negative at the bounce ($\Hu=0$,
$\Hu'>0$) even though it is positive definite, being also given by
$\varphi^{\prime 2}/a^2$. Indeed, it is well known that in order to
have a bouncing period in a FLRW background, one must violate energy
conditions that classical fluids usually do satisfy (as was presented
in Ref.~\cite{ekpN} and whose origin can be in fact traced back to
Ref.~\cite{global}). If one insists on having a flat ${\cal K}=0$
situation with, say, a scalar field alone, one must either use other
equations, or assume the existence of a singularity.

The only way to have a bounce in GR with a well behaved (NEC
preserving) hydrodynamical fluid as the only source of energy momentum
is in the case of a closed, ${\cal K}=1$ universe. In this case
however, as recently discussed~\cite{ppnpn1}, one finds that the
bounce must be followed by an inflationary epoch, thus considerably
lowering the interest of the model as an alternative to inflation. We
shall nevertheless study this case as the only fully self-consistent
possibility.

In the literature~\cite{BF,Hwang}, it was suggested to treat the
bounce as a GR sharp transition, i.e., to assume that the time scale
of the bounce is very short and therefore that the theory has ``no
time'' to deviate too strongly from GR. The continuous and
self-consistent GR model developed here will allow us to test the
validity of these hypothesis, namely that the bounce can be
appropriately approximated by a sharp transition between the slow
contraction phase and the radiation era, and that the curvature
perturbation $\zeta$ is continuous in agreement with the standard
junction conditions.

\begin{figure}[ht]
\includegraphics*[width=8.5cm, angle=0]{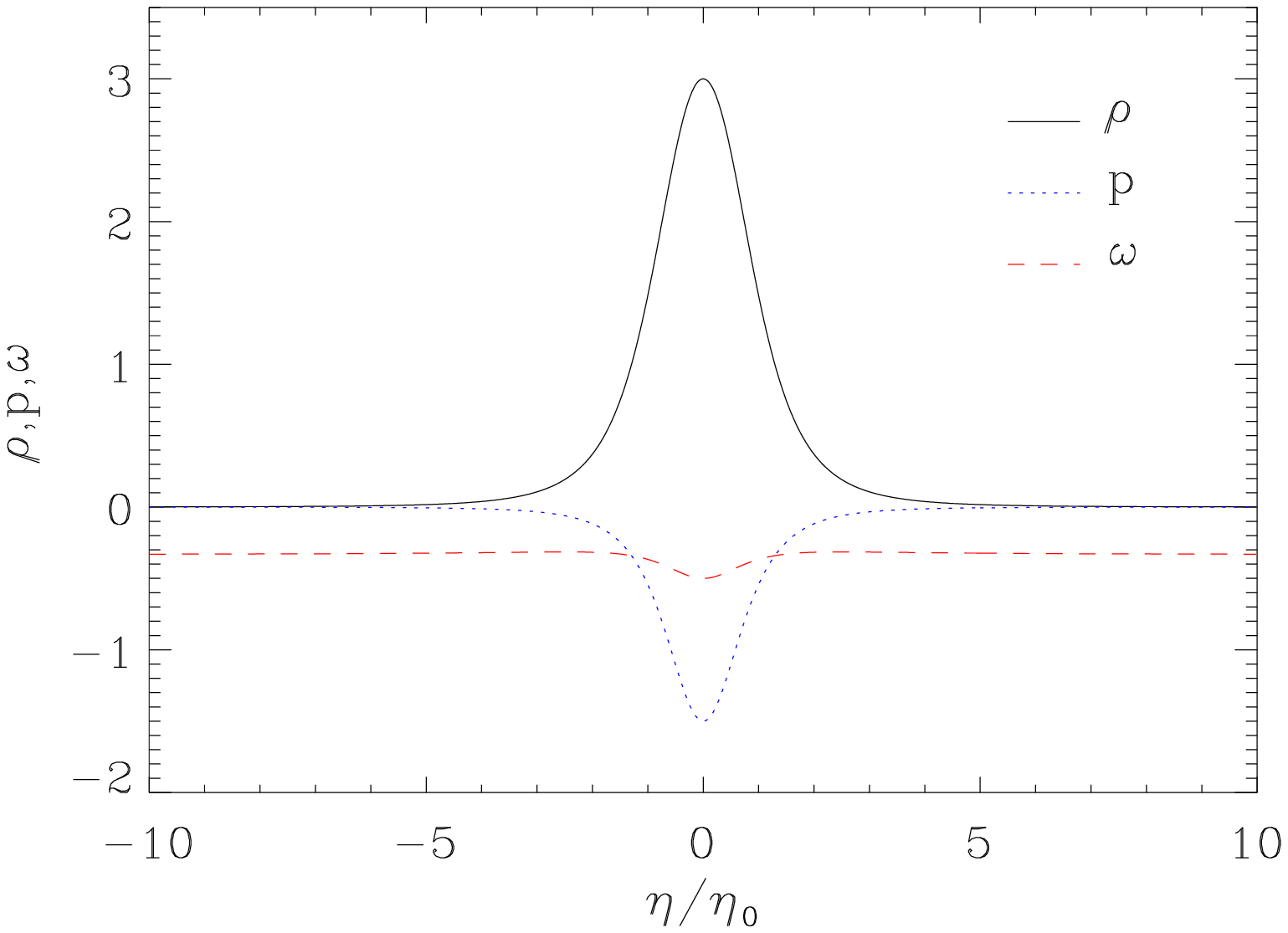}
\includegraphics*[width=8.5cm, angle=0]{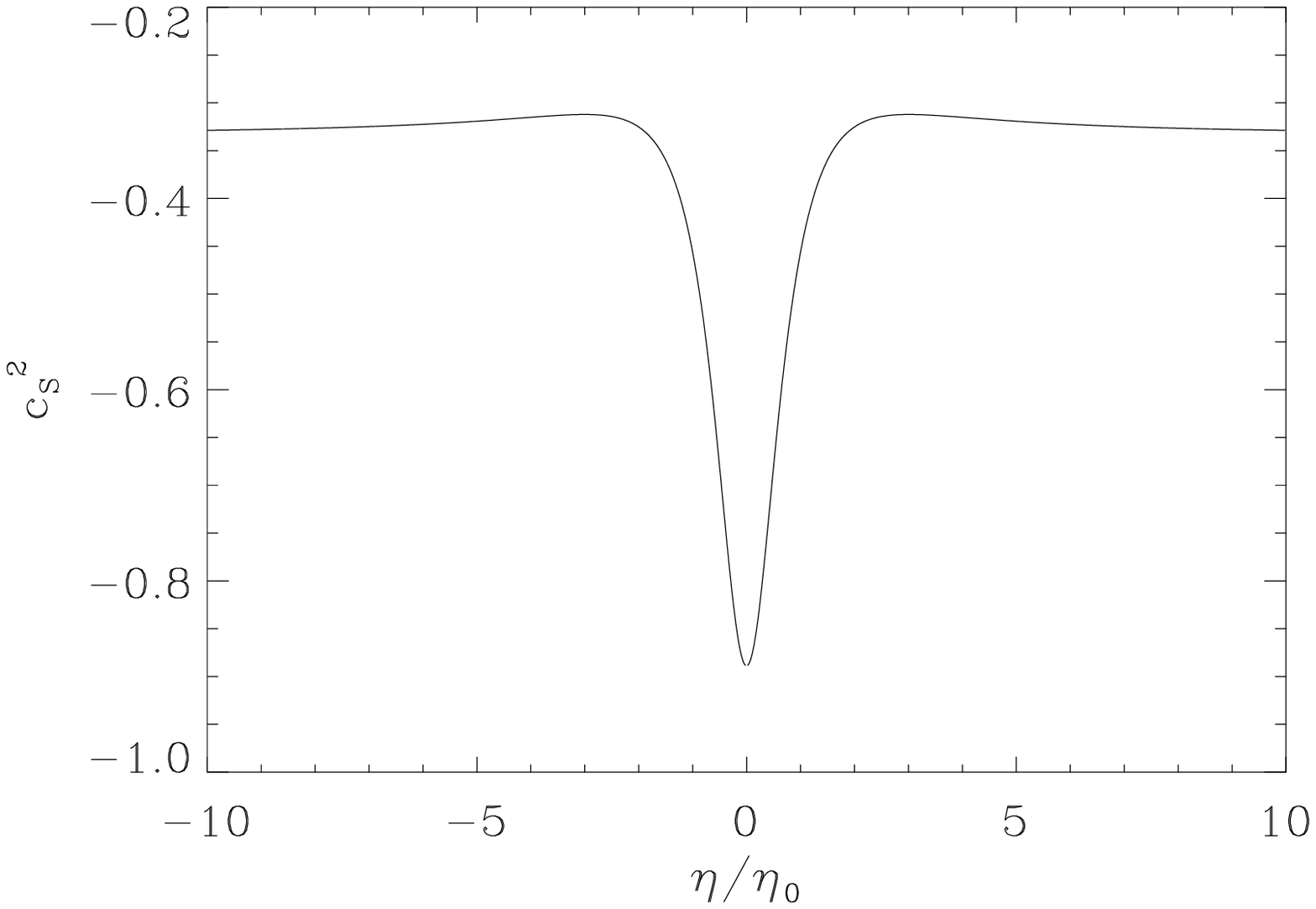}
\caption{Components of the stress energy tensor (top panel) deduced
from Eqs.~(\ref{G00}) and (\ref{Gij}) as functions of the conformal
time $\eta/\eta_0$ for the scale factor~(\ref{scale}). Shown are the
values of the background energy density $\rho$ (full line), the
pressure $p$ (dotted line) and their ratio $\omega$ (dashed line). In
this figure as well as the following, the parameters have been given
the particular values $\ell _0=1$ and $\eta _0=2$. The bottom panel
displays the sound velocity $\cs^2$ as given by Eq.~(\ref{cs2reg}).}
\label{fig:bounce1}
\end{figure}

Let us now turn to the description of the model considered in the
following sections. We choose the behavior of the scale factor
around the bounce to be
\begin{equation}
a(\eta )= \ell _0\biggl[1+\frac{1}{2}\biggl(\frac{\eta }{\eta
_0}\biggr)^2\biggr].\label{scale}
\end{equation}
This choice is reasonable since any function describing a bouncing
scale factor can be approximated by a parabola, at least in the
vicinity of the bounce. Any other choice would thus be equivalent to
this one, and the results one would get, for instance by including
higher order extra terms in Eq.~(\ref{scale}) seen as an expansion,
would be qualitatively unchanged. Such a behavior for the scale
factor results from the presence of an hydrodynamical fluid with an
unusual equation of state. The various physical quantities needed to
describe the bounce such as the energy density, pressure, equation of
state and sound velocity are displayed in Fig.~\ref{fig:bounce1}. In
particular, the sound velocity is given by the relation
\begin{equation}
\cS^2= -\displaystyle{
\frac{4+\eta_0^2+\eta_0^2
\left(\displaystyle{\frac{\eta }{\eta _0}}\right)^2 +
\displaystyle{\frac{\eta_0^2}{4}}
\left(\displaystyle{\frac{\eta }{\eta _0}}\right)^4}
{ 3\left[ {\eta_0^2-1+\left(\displaystyle{\frac{3}{2}}+\eta _0^2\right) 
\left(\displaystyle{\frac{\eta
}{\eta _0}}\right)^2 +\displaystyle{\frac{\eta _0^2}{4}}\left(
\displaystyle{\frac{\eta }{\eta_0}}\right)^4 }\right]}}.\label{cs2reg}
\end{equation}

{}From the figures, one can already see that the bounce is rather
unlikely to be well described by a sharp transition which would
require a finite jump in both $\omega$ and $\cs^2$. One could however
argue at this point that this is due to our approximation for the
scale factor at the bounce: an even scale factor leads to an even
equation of state and therefore to a transition which cannot be
assumed sharp, even if the transition duration goes to zero. 

Another, more important, reason to oppose the sharp transition
treatment of the bounce lies in the following. In the neighborhood of
the bounce we have
\begin{equation}\label{nec}
(\rho +p)(\eta =0) = \frac{2}{\ell _0^2}\biggl(1-\frac{1}{\eta
_0^2}\biggr),\end{equation} which can be easily
interpreted. Restoring the usual units (with $c$ the
velocity of light), the above equation can be rewritten as $\kappa
(\rho + p) = 2/\ell_0^2 [\Ka-\ell_0^2/(c^2 t_0^2)]$, in which we
interpret $\ell_0$ as the curvature scale [$\equiv
a(\eta)/\sqrt{|\Ka|}$] at the bounce, and $t_0=\ell_0\eta_0/c$ is the
physical time taken by light to go across the bounce. It makes sense
that some exotic matter ($\rho+p<0$) is required if the growth of the
universe is faster than light. Similarly, the sound velocity given by
Eq.~(\ref{cs2}) reveals that if $\eta _0\leq 1$, there always is a
point at which $\cS^2$ diverges. This is connected to the violation of
the null energy condition~\cite{ppnpn1} seen in Eq.~(\ref{nec}). This
has of course important consequences with respect to our wish to have
a short duration bounce. In particular, it means that one cannot, in
this framework, investigate the short bounce limit for which
$\eta_0\to 0$. Therefore, we reach the conclusion that the time scale
of the bounce cannot be made arbitrary short if we want to deal only
with well-behaved quantities. This provides another argument against
the sharp transition limit. We shall for now on restrict ourselves to
the case $\eta _0>1$, which is consistent with our choice of setting
$\Ka=1$ in order to avoid unnecessary exotic matter.
\par

Let us now discuss the standard junction conditions. For the
background, matching by brute force the pre- and post-impact phases
(i.e., assuming that the bounce time scale is negligible) means
$[\Hu]\neq 0$ since $\Hu$ has not the same sign before and after the
bounce. On the other hand, the junction conditions applied to the
background demand that $[\Hu]=0$. Therefore, it seems that it is
already impossible to use the standard GR conditions at the background
level, as pointed out in~\cite{Ruth}. This was also the reason why, in
Ref.~\cite{Ruth}, a surface layer pressure term was introduced such as
to allow for a jump in $\Hu$.
\par

Accordingly, let us admit that we can study the perturbative level. It
has been shown in Ref.~\cite{MS} that the well-known cosmological
perturbation matching conditions for ${\cal K}=0$ also holds for
${\cal K}=1$. In this last case, the quantity $\zeta$ is not
conserved, and it is better to work with another quantity $\zBST$ defined by
\begin{widetext}
\begin{eqnarray}
\zBST &\equiv & -\frac{2}{3} \frac{{\cal H}^2}{(1+\omega
)({\cal H}^2+{\cal K})} \biggl\{{\cal H}^{-1}\Phi
'+\biggl[1-\frac{{\cal K}}{{\cal H}^2}
+\frac{1}{3}\biggl(\frac{k}{{\cal H}}\biggr)^2\biggr]\Phi\biggr\}
-\Phi \\ &=& -\frac{{\cal H}^2}{{\cal H}^2+\Ka}\zeta -\frac{k^2}{3\Gamma
{\cal H}^2}\Phi -{\cal K}\biggl(\frac{1}{{\cal H}^2+{\cal
K}}-\frac{1}{\Gamma {\cal H}^2} \biggr)\Phi.\label{zetaBST}
\end{eqnarray}
\end{widetext}
The derivative with respect to conformal time of the previous
quantity reads
\begin{equation}
\frac{1}{{\cal H}}\zBST'= -\frac{2}{3} \frac{{\cal
H}^2}{(1+\omega )({\cal H}^2+{\cal K})}
\biggl[\frac{1}{3}\biggl(\frac{k}{{\cal H}}\biggr)^2 ({\cal
H}^{-1}\Phi '+\Phi )\biggr].\label{zetaBSTprime}
\end{equation}
Therefore, we see that, even if ${\cal K}\neq 0$, $\zBST$ 
is approximately constant on superhorizon scales and this 
property has been used in Refs.~\cite{BF,Hwang}. With the exact 
toy model at our disposal, this can be explicitly tested. 
\par

\begin{figure}
\includegraphics*[width=8.5cm, angle=0]{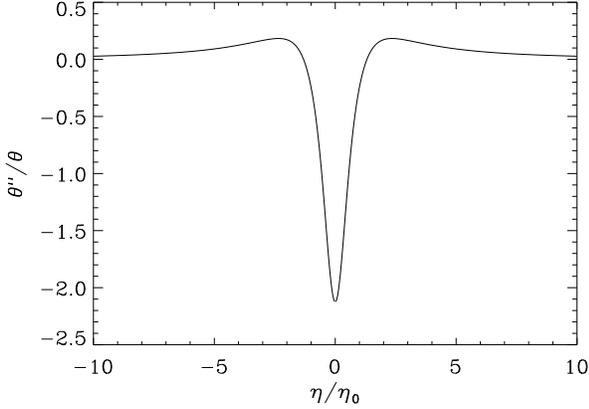}
\caption{The parametric oscillator potential $\theta''/\theta$ [See
Eqs.~(\ref{mutheta}) and (\ref{pot})] as functions of the conformal
time with the same parameter as Fig.~\ref{fig:bounce1}.}
\label{fig:bounce2}
\end{figure}

\begin{figure}[ht]
\includegraphics*[width=8.5cm, angle=0]{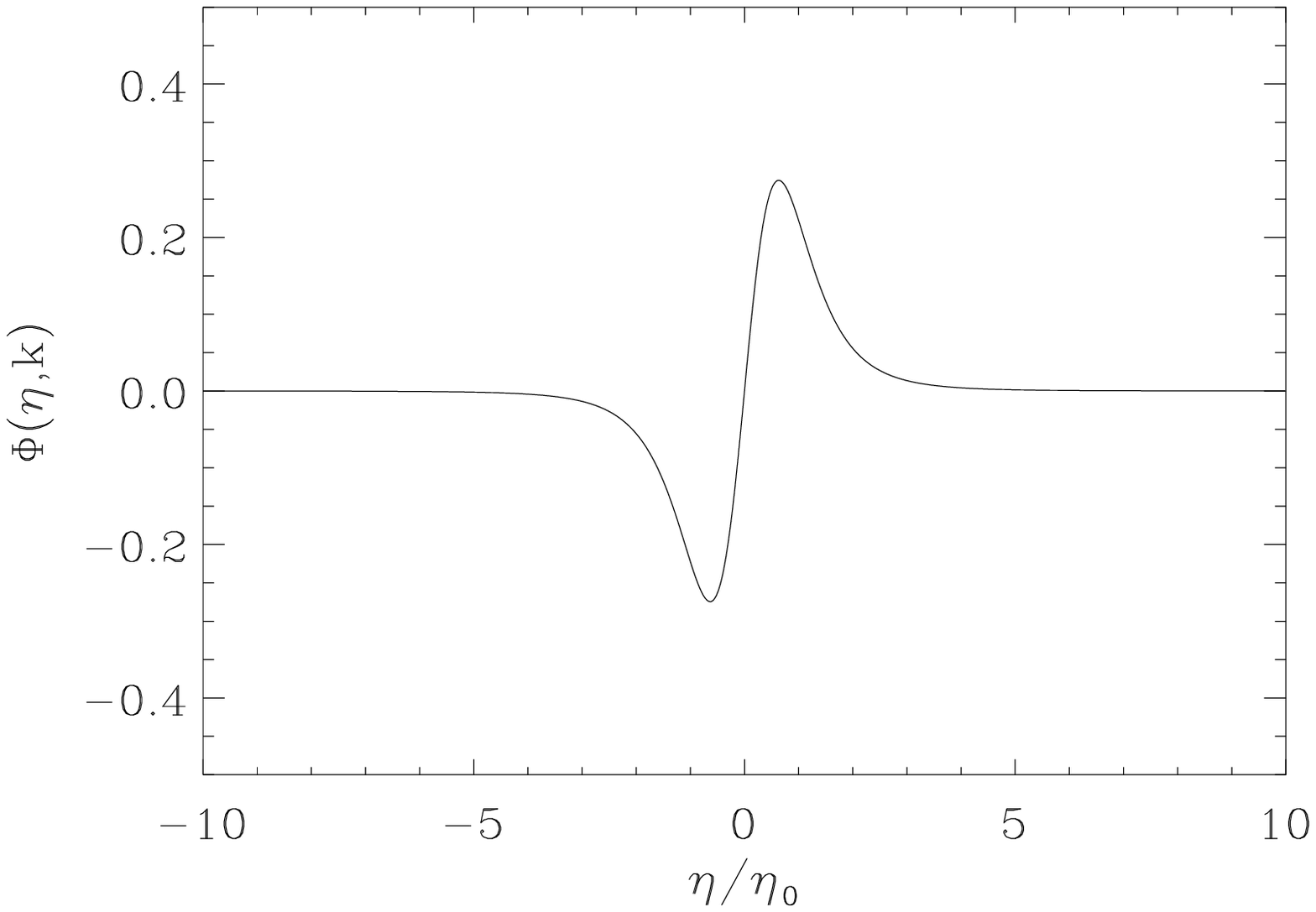}
\includegraphics*[width=8.5cm, angle=0]{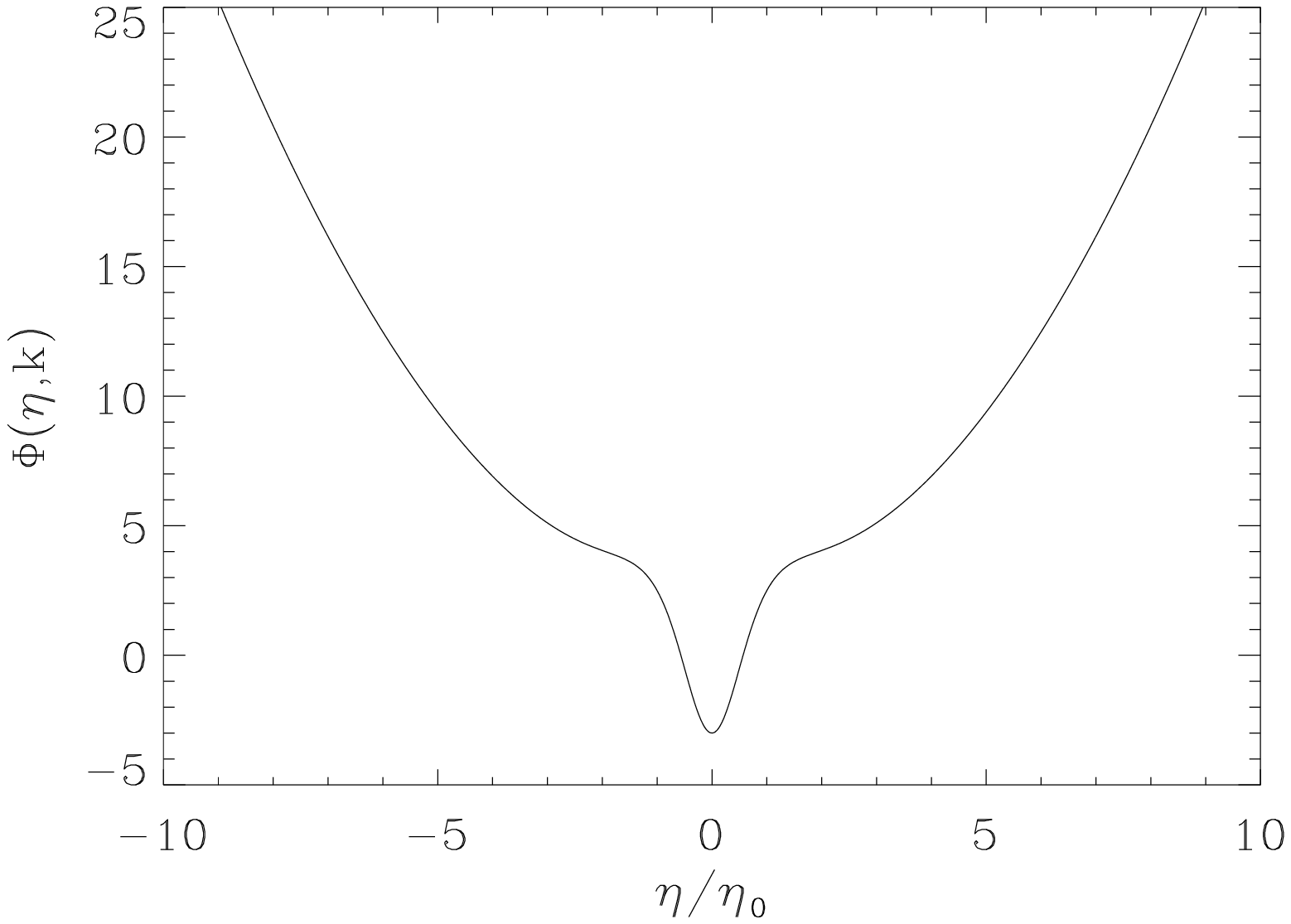}
\caption{Odd (top) and even (bottom) modes of the Bardeen potential
[see Eq.~(\ref{Phioddeven})] through the parabolic bounce as functions
of the conformal time with the same parameter as
Fig.~\ref{fig:bounce1}.}
\label{fig:Phi}
\end{figure}

We now turn to the study of the perturbed quantities. The effective
potential $\theta ''/\theta $, see Eq.~(\ref{mutheta}), for the
scalar perturbations is given in Fig.~\ref{fig:bounce2}.
We assume that there is no entropy production and the equations
governing the evolution of Bardeen potential are Eqs.~(\ref{Bardeen})
and~(\ref{pot}). Two cases must be studied. In the short wavelength
limit, for which $\cS^2 (k^2-\Ka) \gg \theta''/\theta$, $\zBST$ is
clearly not conserved. Therefore, the only case which remains to be
studied is that of long wavelengths. The latter approximation can be
applied if $\cS^2 (k^2-\Ka) \ll \theta'' /\theta$, where $k=1, 2,
\cdots$. This is obviously true for $k=1$, i.e., a mode that cannot be
confused with the background. This is less clear for higher $k$ modes,
and depends on the parameter values. In the short bounce limit
$\eta_0\to 1$ we are interested in, one has $\cS^2 (\eta=0) \sim
-(5/3) (\eta_0^2-1)^{-1} -1/3 +{\cal O}(\eta_0^2-1)$ and
$\theta''/\theta (\eta=0)\sim -(15/2) (\eta_0^2-1)^{-1} +3/2 +{\cal
O}(\eta_0^2-1)$, so that the ratio tends to the fixed value $9/2$. In
this limit, the mode $k=2$ also marginally satisfies the long
wavelength requirement. The approximation breaks down for $k\geq
3$. Since there exists at least one physically meaningfull mode for
which the approximation is valid, we can proceed and use
Eq.~(\ref{Phisol}) for the relevant modes. In the case at hand, the
integral can be performed exactly and the final result can be written
as
\begin{widetext}
\begin{eqnarray}
\Phi (\eta ,k)&=& B_1(k)\frac{3}{2\ell _0^2\eta _0}\frac{\eta
}{\eta _0} \biggl[1+\frac{1}{2}\biggl(\frac{\eta }{\eta
_0}\biggr)^2\biggr]^{-3} +B_2(k)
\biggl[1+\frac{1}{2}\biggl(\frac{\eta }{\eta
_0}\biggr)^2\biggr]^{-3} \biggl[1-\eta _0^2+\frac{1}{2}(4\eta _0^2
+1)\biggl(\frac{\eta }{\eta _0}\biggr)^2
\nonumber \\
& &+\frac{1}{12}(6\eta _0^2+5)\biggl(\frac{\eta }{\eta
_0}\biggr)^4 +\frac{1}{40}(3+4\eta _0^2) \biggl(\frac{\eta }{\eta
_0}\biggr)^6 +\frac{\eta _0^2}{112}\biggl(\frac{\eta }{\eta
_0}\biggr)^8\biggr].\label{Phioddeven}
\end{eqnarray}
\end{widetext}
As expected one branch is even and the other is odd. One also
sees that if $\eta _0=1$ the minimum of the even branch is zero.
This makes sense since in this case $\rho+p=0$, i.e., the de Sitter
equation of state for which it is known that $\Phi =0$. The two
branches are plotted in Fig.~\ref{fig:Phi}.

\begin{figure}[ht]
\includegraphics*[width=8.5cm, angle=0]{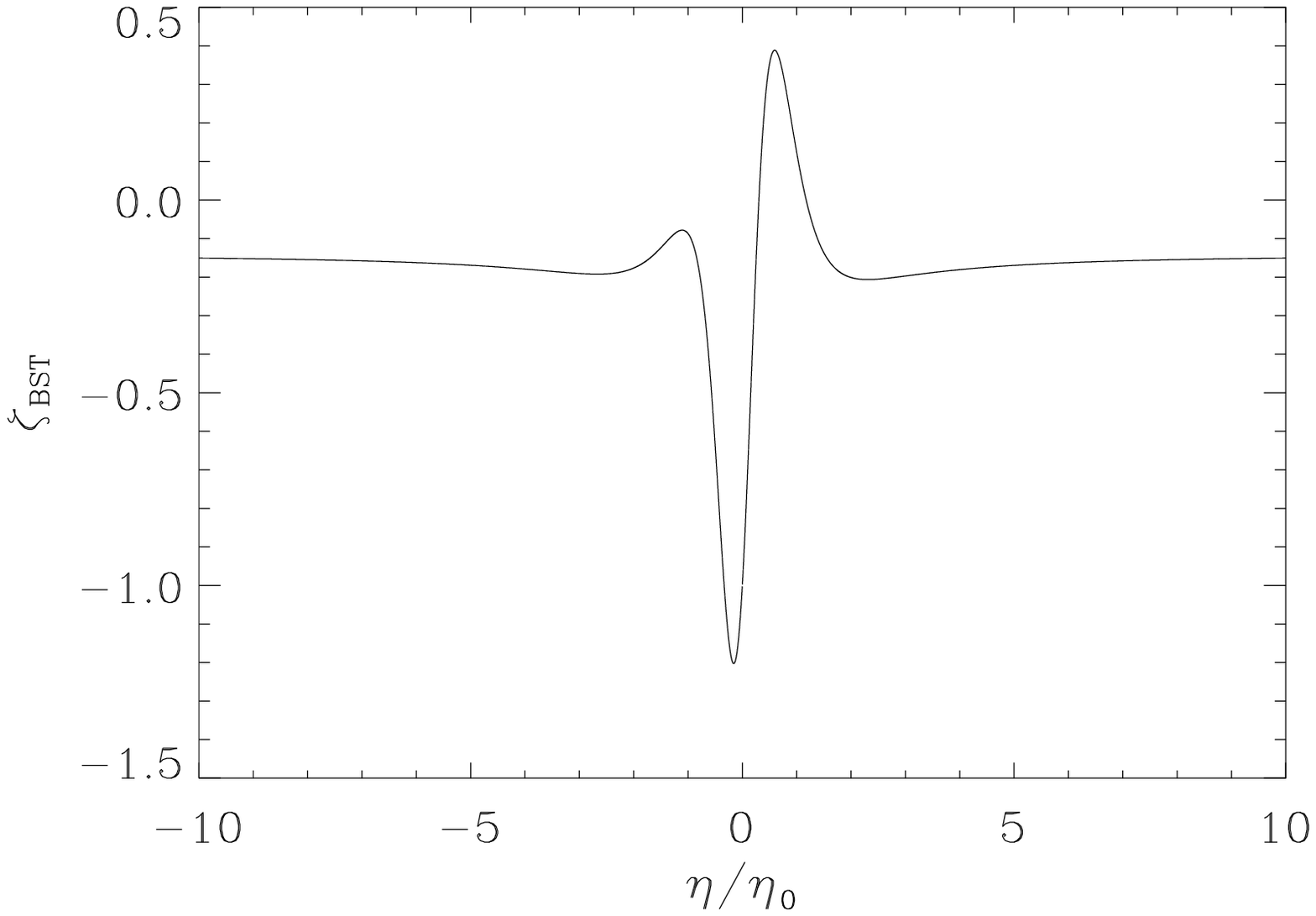}
\includegraphics*[width=8.5cm, angle=0]{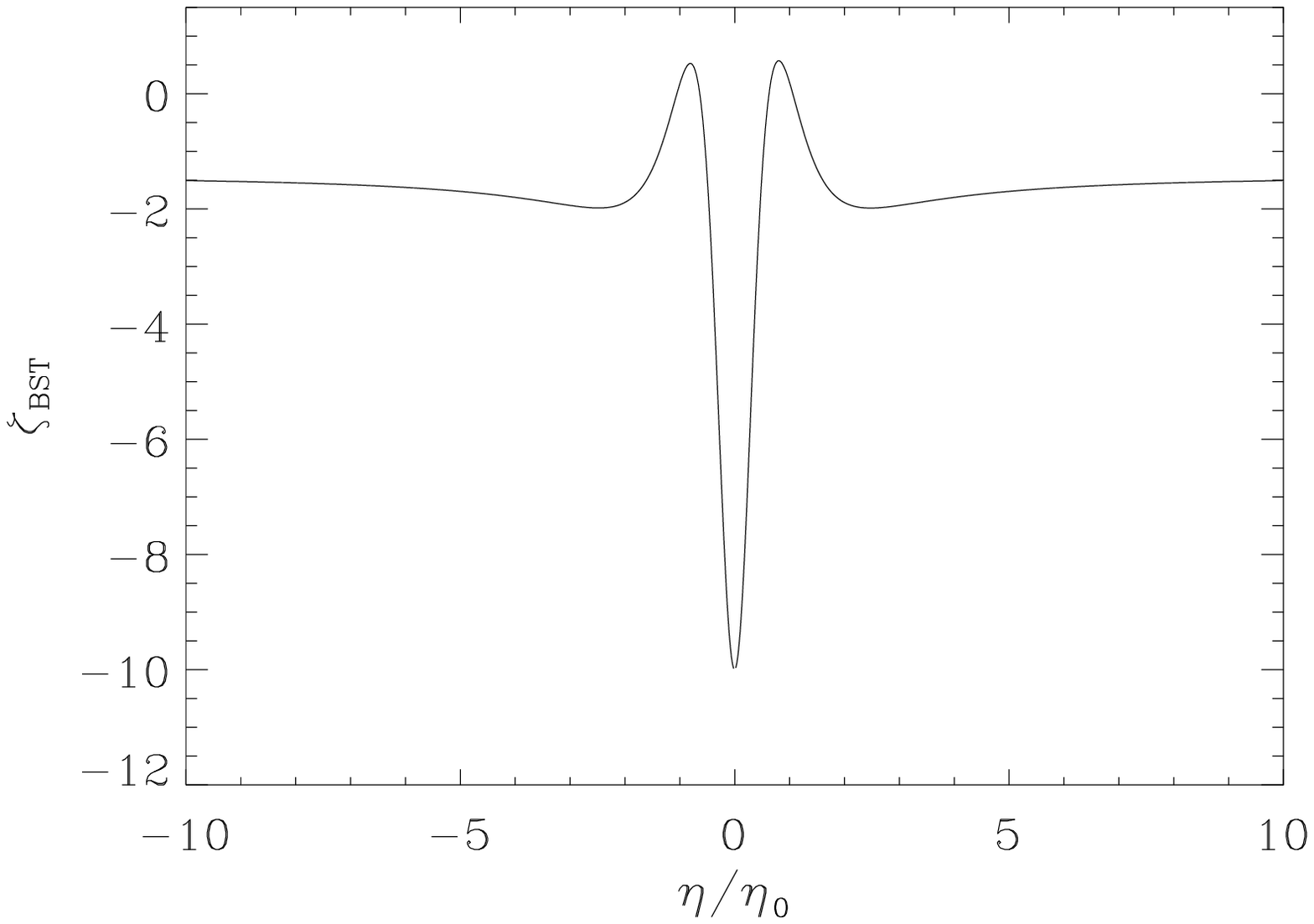}
\caption{Two typical configurations for the supposedly conserved
quantity $\zBST$ as functions of the conformal time with
the same parameter as on the previous figures. Top panel: $B_1=10$
and $B_2=1$. Bottom panel: $B_1=1$ and $B_2=10$.} \label{fig:zeta}
\end{figure}

We are now in the position where we can estimate $\zBST$ for these
long wavelength modes. Expanding Eq.~(\ref{zetaBST}) to leading order,
one finds
\begin{equation}
\label{approxzetaBST}
\zBST\simeq -\frac{{\cal H}^2}{{\cal H}^2+{\cal K}}B_2
-{\cal K}\biggl(\frac{1}{{\cal H}^2+{\cal K}}-\frac{1}{\Gamma
{\cal H}^2} \biggr)\Phi +{\cal O}(k^2).
\end{equation}
In the above relation, only the $B_2$--mode of $\zeta $ appears in the
first term since the $B_1$--term is of order $k^2B_1(k)$, see
Eq.~(\ref{zetasol}), whereas the corresponding term in the Bardeen
potential is of order $B_1(k)$. Far from the bounce, when $\eta /\eta
_0\gg 1$, the first term in Eq.~(\ref{approxzetaBST}) tends to zero,
while the second goes to $-B_2/7$ on both sides, as can be explicitly
checked in Figs.~\ref{fig:zeta}. Therefore, if we consider a long time
interval $\Delta \eta /\eta _0\gg 1$, this quantity seems to be indeed
constant.  On the other hand, close to the bounce, over typical bounce
time scales, i.e., $\Delta \eta /\eta _0\simeq 1$, one clearly sees in
Figs.~\ref{fig:zeta} that the quantity $\zBST$ is not a constant. This
is due to the fact that, during the bounce, the ``growing'' and
``decaying'' modes are of the same order of magnitude as shown in
Fig.~\ref{fig:zeta}.

\par To conclude this section, let us summarize what we have learned
from the simple toy model used here: (i) we have seen that the
standard GR junction conditions applied to the background are not
consistent with a bounce since $\Hu$ has not the same sign before and
after the bounce~\cite{Ruth}, (ii) we have shown that the equation of
state does not jump at the bounce, (iii) we have found that a bounce
cannot be made arbitrary short without violating the null energy
condition~\cite{ppnpn1} and finally (iv) we have noticed that the
quantity $\zBST$ is not constant on the typical bounce time
scale (recall that beyond the bounce epoch, our model looses its
meaning and should be matched to another era).

\section{A test scalar field in the ekpyrotic universe}
\label{sec:scalar}

Let us now turn to the last case for which one can explicitly
calculate the various physically meaningful quantities during a
regular bounce. We now assume that a bounce took place, and consider
perturbations of a test scalar field in that background. In this case,
we do not need to specify the origin of the scale factor. As in
section~\ref{sec:hydro}, we question the conservation of what we know
is conserved in sharp transitions. We can also regard the
perturbations studied in this section as gravitational waves, provided
then that $\Ka=0$ and that somehow the necessary modification of GR is
negligible on the tensor part of the perturbations.

The equation of a test scalar field in a spatially FLRW spacetime
with a scale factor given by the previous expression is
\begin{equation}
\label{eom2} \mu '' + \biggl(k^2- {{a''} \over a}\biggr)\mu \, =
\, 0 \, .
\end{equation}
As mentioned above, this is also the equation of motion of
gravitational waves in GR if ${\cal K}=0$. The solution of this
equation possesses two regimes determined by the relative contribution
of the two terms $k^2$ and $a''/a$. The transition time $\eta _{\rm
j}(k)$ is defined by $k^2=a''/a$. In the case of a parabolic scale
factor, one has
\begin{equation}
\frac{a''}{a}=\frac{1}{\eta _0^2} \biggl[1+\frac{1}{2}
\biggl(\displaystyle{\frac{\eta }{\eta _0}}\biggr)^2\biggr]^{-1}.
\end{equation}
The maximum of the quantity $a''/a$ is $1/\eta _0^2$ and define
the only characteristic scale of the problem, i.e., $k_{\rm
max}=1/\eta _0$. Let us define the parameter $\epsilon$ by
$\epsilon \equiv k/k_{\rm max}$, then one has
\begin{equation}
\eta _{\rm j}(k)=\pm \frac{\eta _0}{\epsilon }\sqrt{2(1-\epsilon
^2)}.
\end{equation}
The next step is to solve the equation of motion. From the above
considerations, we see that there are three different regions. In
the first region where $\eta <-\eta _{\rm j}(k)$, we only
consider positive frequency modes and we have
\begin{equation}
\mu _{\rm I}(\eta )=\frac{1}{\sqrt{2k}}\exp [-i k (\eta -\eta _{\rm
i})]
\end{equation}
where $\eta _{\rm i}$ is an arbitrary initial time. In the second
region, where $-\eta _{\rm j}(k)<\eta <\eta _{\rm j}(k)$, the
solution is given by
\begin{equation}
\mu _{\rm II}(\eta )= B_1 a(\eta )+B_2a(\eta )\int _0^{\eta }
\frac{{\rm d}\tau }{a^2({\tau })}.
\end{equation}
The lower bound of the integral is {\it a priori} arbitrary.
However, it is very convenient to take it equal to zero because
in this case the second branch becomes odd whereas the first one
(i.e., the scale factor) is even. Then, it is easy to show that
\begin{equation}
\int _0^{\eta }\frac{{\rm d}\tau }{a^2({\tau })}= \frac{\eta
_0}{2\ell _0}\frac{1}{a(\eta )} \biggl[\frac{\eta }{\eta
_0}+\frac{\sqrt{2}}{\ell _0}a(\eta ) \tan ^{-1}\biggl(\frac{\eta
}{\sqrt{2}\eta _0}\biggr)\biggr].\label{conssol}
\end{equation}
Finally, the solution in the third region where $\eta >\eta _{\rm
j}(n)$ can be written as
\begin{equation}
\mu _{\rm III}(\eta )= \frac{C_1}{\sqrt{2k}}\exp (-ik\eta )
+\frac{C_2}{\sqrt{2k}}\exp (+ik\eta ).
\end{equation}

\begin{figure}[t]
\includegraphics*[width=8.5cm, angle=0]{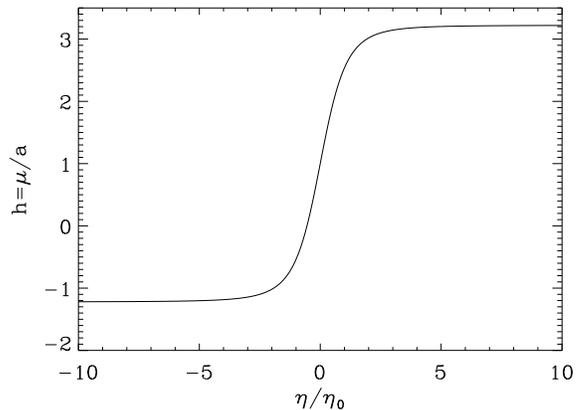}
\caption{The expectedly conserved quantity $\mu/a$ for a spectator
scalar field (or gravitational wave production) in a bouncing
universe, i.e., Eq.~(\ref{conssol}), with parameter fixed at $\ell
_0=1$, $\eta _0=2.$, $B_1=1$ and $B_2=1$. It is clear on this figure
that this quantity is not actually conserved through the bounce.}
\label{fig:zetagw}
\end{figure}

Fig.~\ref{fig:zetagw} shows the solution (\ref{conssol}) as a function
of the conformal time through the bounce. In the usual situation, 
the quantity $h=\mu /a$ is conserved because only the growing 
mode $a$ plays a role. In the present situation, it is clear that 
the usually conserved quantity is not actually conserved. The reason 
is that through the bounce the odd mode (which is the decaying 
mode in general) $a\int ^{\rm \eta }{\rm d}\tau /a^2$ now plays 
a crucial role. Therefore if we match the bounce epoch to other 
eras before and after the impact, we see that the usual conservation 
cannot be used. 
\par

Since the conservation law cannot be utilized, one has to 
perform the calculation explicitly. Therefore, the goal is now to 
calculate the coefficients $C_1$ and $C_2$.  Using
the continuity of the mode function $\mu $ and of its derivative, we
find
\begin{widetext}
\begin{eqnarray}
\label{coeffC1} C_1(k) &=& 
\frac{\ex ^{i k (2 \eta _{\rm j}+\eta _{\rm i})}}{2 i k}
\biggl\{-[g'+i k g](-\eta _{\rm j}) [f'-i k f](\eta _{\rm
j})+[f'+i k f](-\eta _{\rm j})
[g'-i k g](\eta _{\rm j})\biggr\}, \\
\label{coeffC2} C_2(k) &=& -\frac{\ex ^{i k\eta _{\rm i}}}{2 i k}
\biggl\{-[g'+i k g](-\eta _{\rm j}) [f'+ i k f](\eta _{\rm
j})+[f'+i k f](-\eta _{\rm j}) [g'+i k g](\eta _{\rm j})\biggr\},
\end{eqnarray}
where we have used the short-hand notation $f\equiv a(\eta )$ and
$g\equiv a(\eta )\int ^{\eta }_0{\rm d}\tau /a^2(\tau )$. The
final result is given by Eqs.~(\ref{coeffC1}), (\ref{coeffC2})
where all the functions are explicitly known except the function
$\eta _{\rm j}=\eta _{\rm j}(k)$. Expanding everything in terms
of the small parameter $\epsilon$ one finds
\begin{equation}
C_1(k)=\frac{\pi }{4(k\eta
_0)^3}\ex ^{2i\sqrt{2}}(4+i\sqrt{2})+{\cal O}(\epsilon ^{-2}), 
\quad
C_2(k)=-\frac{3i\pi }{2\sqrt{2}(k\eta _0)^3}
+{\cal O}(\epsilon ^{-2}).
\label{spec1}
\end{equation}
\end{widetext}
The spectrum is defined by the following expression
\begin{equation}
k^3P(k)=\frac{k^3}{2\pi ^2}\biggl\vert \frac{\mu (+\eta _{\rm
j})}{a(+\eta _{\rm j})}\biggr \vert ^2.
\end{equation}
Since the coefficients $C_1$ and $C_2$ are of the same order in $k$,
as shown in Eqs.~(\ref{spec1}), the power spectrum $k^3P(k)$ will take
the form of an overall power-law amplitude times an oscillatory
function in $k$. If we parameterize the overall amplitude as
$k^{\nS-1}$ for $k\ll k_{\rm max}$, one has $\nS=1$, i.e, a
scale-invariant spectrum. In the other limit $k\gg k_{\rm max}$ where
Eqs.~(\ref{coeffC1}), (\ref{coeffC2}) cannot be used, the result is
obvious since the $k^2$ term always dominates the effective potential
$a''/a$ in Eq.~(\ref{eom2}): it is $\nS=3$. We also note that when
$\eta _0 \rightarrow 0$, the spectrum blows up and this provides
another argument against a sharp bounce.
\par

In order to test the dependence of the spectrum in the precise
form of the bounce, it is interesting to calculate the spectrum
for another scale factor. We choose
\begin{equation}
a(\eta )=\ell _0\sqrt{1+\biggl(\frac{\eta }{\eta
_0}\biggr)^2}.\label{aroot} 
\end{equation}
This case is treated in Ref.~\cite{ppnpn2}. Let us notice that when
$\eta $ is small (i.e., close to the bounce) the previous equation
reduces to Eq.~(\ref{scale}) as expected. Then straightforward 
calculations lead to, still for $k\ll k_{\rm max}$,
\begin{equation}
C_1(k)=\frac{i\pi }{2k\eta _0}+{\cal O}(\epsilon ^0), \quad
C_2(k)=-\frac{i\pi }{2k\eta _0}+{\cal O}(\epsilon ^0) .
\end{equation}
In this case, since $C_1=-C_2$ at leading order, the oscillatory part
of the power spectrum contributes in a non trivial way to the spectral
index of the overall amplitude. We find $\nS=3$ which, by comparison
with the spectrum obtained in Eq.~(\ref{spec1}), shows that this
spectrum is very strongly dependent on the actual shape of the bounce.
This was to be expected since the bounce has already been shown not to
be a sharp transition.  \par

The conclusion of this section is that the spectrum of gravitational
waves (if we accept the trick that, in a bouncing universe, the
spectrum of a free scalar field can be a good approximation of the
actual gravitational waves spectrum) is in general more complicated
than in the inflationary case. A first feature is that there exists 
a prefered scale the magnitude of which depends on the details 
of the model. A second property is that, generically, the power 
spectrum acquires superimposed oscillations due to the 
fact that, at last horizon entry, the two branches contribute 
equally. Therefore, the shape of the spectrum crucially
depends on the details of the model.

\section{Conclusions}

The conclusions that can be drawn from this work is that it seems
impossible to apply any known and well motivated criterion to pass
through a bounce, whether regular or singular, in a model independent
way as all quantities of interest explicitly depend on the details of
the underlying model. The ekpyrotic model, although a potentially
interesting alternative to the inflationary paradigm, does pass
through such a bounce. Therefore, if one really wants to calculate
the spectrum in the ekpyrotic universe then it seems necessary, first,
to consider a situation where there is no divergence and, second, to
provide us with the actual (maybe five-dimensional) equations of
motion during the bounce, knowing that these equations cannot be those
of GR.

\acknowledgments We would like to thank R.~Brandenberger and
G.~Veneziano for careful reading of the manuscript and for various
illuminating comments. It is also a pleasure to thank A.~Buonanno,
J.~Hwang, D.~Lyth and N.~Turok for interesting discussions and/or
remarks.

\end{document}